\documentclass[journal]{IEEEtran}

\newcommand{\E}{\mathbb{E}}

\newcommand{\NMSE}{{\mbox{NMSE}}}

\newcommand{\DEF}{\triangleq}

\newcommand{\bfy}{{\bf y}}

\newcommand{\bfB}{{\bf B}}
\newcommand{\bfQ}{{\bf Q}}
\newcommand{\bfC}{{\bf C}}

\newcommand{\bfx}{{\bf x}}

\newcommand{\bff}{{\bf f}}
\newcommand{\bfG}{{\bf G}}

\newcommand{\bfe}{{\bf e}}
\newcommand{\bfE}{{\bf E}}

\newcommand{\bfr}{{\bf r}}

\newcommand{\bfv}{{\bf v}}
\newcommand{\bfu}{{\bf u}}

\newcommand{\bfxt}{\widetilde{{\bf x}}}
\newcommand{\bfA}{{\bf A}}

\newcommand{\bfw}{{\bf w}}

\newcommand{\bfU}{{\bf U}}
\newcommand{\bfUb}{\overline {\bf U}}
\newcommand{\ub}{\overline u}

\newcommand{\bfUbb}{\overline {\overline {\bf U}}}
\newcommand{\ubb}{\overline {\overline {u}}}

\newcommand{\bfXt}{\widetilde{{\bf X}}}

\newcommand{\bfV}{{\bf V}}

\newcommand{\bfnoll}{{\bf 0}}
\newcommand{\bfI}{{\bf I}}
\newcommand{\bfK}{{\bf K}}
\newcommand{\bfL}{{\bf L}}
\newcommand{\Px}{P_{x}}

\newcommand{\vect}[1]{\mathbf{#1}}

\usepackage{color}
\usepackage{amsfonts}
\usepackage{amssymb}
\usepackage{cite}
\usepackage{color}
\definecolor{blue}{rgb}{0,0,0}
\ifCLASSINFOpdf
 
\else
  
\fi
\usepackage{amsmath}
\usepackage{stfloats}
\usepackage{graphicx}
\usepackage{psfrag}
\newtheorem{thm}{Theorem}
\newtheorem{lemma}{Lemma}

\begin{document}

\title{Impact of Backward Crosstalk in $2 \times 2$ MIMO Transmitters on NMSE and Spectral Efficiency}

\author{Peter H\"andel,~\IEEEmembership{Senior Member,~IEEE,} \"Ozlem Tu\u{g}fe Demir,~\IEEEmembership{Member,~IEEE,} Emil Bj\"ornson,~\IEEEmembership{Senior Member,~IEEE,} and Daniel R\"onnow,~\IEEEmembership{Member,~IEEE}
\thanks{P.\ H\"andel (deceased) was with the Department of Information Science and Engineering, KTH Royal Institute of Technology, 114 28 Stockholm, Sweden. \"O. T. Demir and E. Bj\"ornson are with the Department of Electrical Engineering
	(ISY), Linköping University, 581 83 Linköping, Sweden (e-mail: ozlem.tugfe.demir@liu.se, emil.bjornson@liu.se). D.\ R\"onnow is with the Department of Electronics, Mathematics and Natural Sciences, University of G\"avle, 801 76 G\"avle, Sweden (e-mail: daniel.ronnow@hig.se).}
\thanks{The work of \"O. T. Demir and E. Bj\"ornson  was partially supported by ELLIIT and the Wallenberg AI, Autonomous Systems and Software Program (WASP) funded by the Knut and Alice Wallenberg Foundation. The work of D.\ R\"onnow was partially financed by the European Commission within the European Regional Development Fund, the Swedish Agency for Economic and Regional Growth, and Region G\"avleborg.}}

\maketitle

\begin{abstract}
We consider backward crosstalk in $2\times 2$ transmitters, which is caused by crosstalk from the outputs of the transmitter to the inputs or by the combination of output crosstalk and impedance mismatch. We analyze its impact via feedback networks together with third-order power amplifier non-linearities. We utilize the Bussgang decomposition to express the distorted output signals of the transmitter as a linear transformation of the input plus uncorrelated distortion. The normalized mean-square errors (NMSEs) between the distorted and desired amplified signals are expressed analytically and the optimal closed-form power back-off that minimizes the worst NMSE of the two branches is derived. In the second part of the paper, an achievable spectral efficiency (SE) is presented for the communication from this ``dirty'' transmitter to another single-antenna receiver. The SE-maximizing precoder is optimally found by exploiting the hardware characteristics. Furthermore, the optimal power back-off is analyzed for two sub-optimal precoders, which either do not exploit any hardware knowledge or only partial knowledge. The simulation results show that the performance of these sub-optimal precoders is close-to-optimal. We also discuss how the analysis in this paper can be extended to transmitters with an arbitrary number of antenna branches.
\end{abstract}


\IEEEpeerreviewmaketitle

\section{Introduction}

Techniques to handle transmitter imperfections, including
crosstalk between the transmitter branches, nonlinearity of the power amplifiers, mixer imbalance, and leakage are of utmost importance for future wireless systems, and is an active field of research \cite{MIMO1,MIMO2,r2,r1,revised2,jessica4}. Transmitter imperfections can be combatted to increase the communication performance or appear as a side-effect of simplified design or implementation.

To complement the derivation of novel methods to combat the transmitter imperfections,
there has been a recent focus on improving the understanding of the imperfections in single-input-single-output (SISO) and multiple-input-multiple-output (MIMO) transmitters under orthogonal frequency-division multiplexing (OFDM) signals.
Recent works include
\cite{jessicaLetter,jessica3} that study different aspects of the normalized mean squared error (NMSE) for a SISO transmitter subject to ideal digital predistortion. A lower bound on the NMSE is derived in \cite{jessicaLetter}. Additional results to those in \cite{jessicaLetter} are provided in \cite{jessica3}, where simple-to-interpret closed-form formulas for the NMSE in different regions of power amplifier compression are obtained.
The same methodology is used  in  \cite{jessica5} to analyze the joint effect of the mixer and power amplifier imperfections in a SISO transmitter, where it is shown that the performance at the NMSE-minimizing power back-off is limited by the imperfections in the IQ-modulator.

A MIMO transmitter has additional artifacts compared with a SISO transmitter, including leakage/crosstalk between the transmitter branches or antennas, that negatively influence its performance  \cite{MIMO1}.  $2 \times 2$ MIMO transmitter structures have been proposed for IEEE 802.11 \cite{ny1,WiFi},  long term evolution (LTE) \cite{LTE}, and 79 GHz radar \cite{Radar}. Several works also focused on the design of digital predistorters to compensate the adverse effects of crosstalk and power amplifier distortions for wide-band code-divison multiple access (WCDMA) and  Worldwide Interoperability for Microwave Access (WiMAX) $2 \times 2$ transmitters \cite{MIMO1,MIMO2,revised2}. $2\times 2$ transmitters are of great importance in studying the performance degradation resulting from crosstalk due to the adjacent antenna branches. The motivation behind this is that the direct coupling between non-adjacent antennas becomes smaller as the distance between the antenna units increases \cite{jessica4}. In this paper, we mainly consider a $2\times 2$ MIMO transmitter, which enables us to follow a completely analytical approach and explore the joint impact of backward crosstalk and power amplifier non-linearities on different parts of a communication system.

In \cite{PHDR1}, a first study of the power amplifier compression distortion and effects of leakage between the branches in a $2 \times 2$ MIMO transmitter is presented, where an analytical expression for the transmitter NMSE is presented for a transmitter subject to crosstalk between the input branches and between the output branches, respectively. Dirty transmitter analysis in the massive MIMO scenario is an identified active area of research \cite{Emil2,Emilny}. Note that these works do not consider crosstalk. The properties of an $M \times M$ transmitter with crosstalk is the subject of \cite{jessica4}, including the asymptotic  massive MIMO regime where $M \rightarrow \infty$. 

The previous works \cite{jessicaLetter,jessica3,jessica5,PHDR1,jessica4} all utilize the classical Bussgang decomposition  \cite{Bussgang} to provide an understanding of the transmitter performance. Despite being theoretical in nature, the Bussgang decomposition has been verified experimentally in both SISO \cite{jessicaLetter} and MIMO  \cite{Nima} scenarios.  It is here emphasized that the main purpose with employing the considered approach  is in the understanding of the transmitter imperfections, including balancing the selection of mixers and transmitters, and effects of coupling between the branches of a transmitter.

\subsection{Contributions}

Almost all of the above-mentioned works consider either linear or non-linear crosstalk that are both modeled by feed-forward connections between the antenna branches \cite{MIMO1, MIMO2, PHDR1, jessica4}.
One transmitter imperfection that has been overlooked in the majority of previous work is the so-called \emph{backward crosstalk} between the MIMO transmitter branches. Backward crosstalk from one amplifier's  output to another's input occurs when there is leakage between transmission lines. A phenomenon with similar effects occur when  there is crosstalk between the outputs of two nonlinear amplifiers that are mismatched \cite{r2,Back1}. Even if the power leakage is small relative to the output power, it can have a large impact since the inputs to the power amplifiers are also small.
For example, if the amplification gain is 20 dB, then a 1\% leakage will result in a crosstalk distortion that is equally strong as the input. The crosstalk appears when the transmitter branches (transmission lines) are physically close and, thus, the issue will likely be larger in future digital mmWave transceivers where many branches must be squeezed into a small circuit.

Different from the existing works \cite{PHDR1,jessica4} that consider forward crosstalk, we consider backward crosstalk, which can be modeled by a feedback network and, hence, is analytically more challenging. Some approximations are introduced along the way to obtain analytically tractable and insightful results. In the simulations, we validate these approximations.

In \cite{Back1}, models for digital predistortion of transmitters under backward crosstalk were proposed and their performance was evaluated in laboratory experiments. \cite{Back1} uses a generalized memory polynomials and does not exploit the Bussgang decomposition to obtain closed-form expressions for the NMSE. In this paper, we provide a deeper level of understanding  of the backward crosstalk by employing the discussed Bussgang decomposition. Explicitly, the paper considers the performance of a $2 \times 2$ MIMO transmitter subject to backward crosstalk, by aid of a fully analytical approach leading to a closed-form expression for the transmitter NMSE, as function of the transmitter imperfections. Transmitter NMSE is one of the common figure-of-merits, which is adapted for studying the effect of hardware impairments \cite{jessicaLetter,jessica3, jessica5}. Different from the existing literature that exploits Bussgang decomposition for modeling the joint distortion caused by crosstalk and power amplifiers, we also consider the spectral efficiency (SE) in data transmission to a single-antenna receiver. The closed-form expressions are used to obtain the optimal power back-off to minimize the maximum of NMSE. Different from the existing literature which exploits Bussgang decomposition for crosstalk impairments, the optimal precoder is derived, which maximizes the SE in data transmission. The optimal input reference power is also found for the conventional maximum ratio transmission (MRT) that is a sub-optimal precoder under backward crosstalk and power amplifier non-linearities. In addition, the SE of another sub-optimal precoder that exploits the hardware impairments to maximize the desired signal strength is analyzed. Lastly, we include a discussion section how the derived closed-form results can be extended to transmitters with an arbitrary number of antennas.
The closed-form expressions and the optimal results are expected to provide the academia and practitioners with a deeper insight into transmitter performance.

\subsection{Outline}

The paper is organized as follows. In Section~\ref{model}, a $2 \times 2$ dirty MIMO transmitter with backward crosstalk is modeled and its properties are analyzed. The model is used in Section~\ref{nmse} to analyze the NMSE at the transmitter output and determine the power back-off for  minimizing the maximum of NMSE of two branches. Then, in Section~\ref{se}, an SE expression of a point-to-point communication system with a single-antenna receiver is derived under backward crosstalk impairment at the transmitter. The optimal precoding vector which maximizes the SE is found analytically. Furthermore, the optimization of the input reference power of the two sub-optimal precoders is considered. Finally, numerical simulations and how to extend the derived results to more than 2-antenna transmitters are included in Section~\ref{numerical} and Section~\ref{extension}. The conclusions are drawn in Section~\ref{conclusion}.  

\textbf{Reproducible research:} All the simulation results can be reproduced using the Matlab code and data files available at: https://github.com/emilbjornson/backward-crosstalk

\textbf{Notation:} $(\cdot)^T$ and $(\cdot)^H$ denote the transpose and Hermitian transpose of a vector, respectively. $\bfI$ is the identity matrix of an appropriate size and $\triangleq$ denotes a definition. $\E[\cdot]$ denotes statistical expectation, while the multivariate circular symmetric complex distribution with covariance matrix ${\bf C}$ is denoted $\mathcal{N}_{\mathbb{C}}({\bf 0},{\bf C})$.

\section{Transmitter Model and Analysis \label{model}}

\begin{figure}[t]
\centering
\includegraphics[width=1\linewidth]{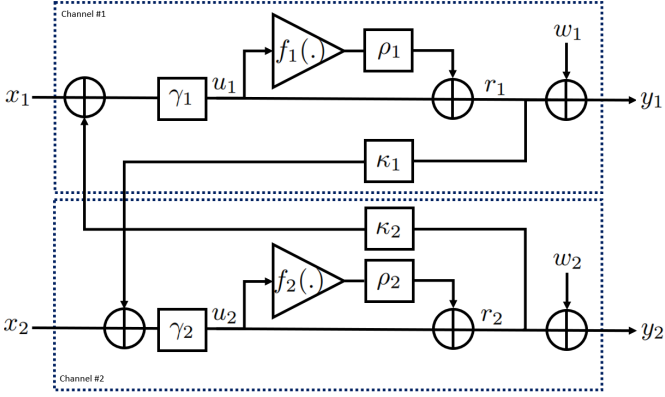} 
\caption{A behavioral model of a $2\times 2$ MIMO transmitter with third-order polynomial nonlinearities $f_\ell(\cdot)$ with compression parameter $\rho_\ell$, subject to backward crosstalk via $\kappa_\ell$ and thermal noise $w_\ell$.} \label{figdirty} 
\end{figure}

In this section, we derive a behavioral model of the $2\times2$ MIMO transmitter with backward crosstalk shown in Fig.~\ref{figdirty}\footnote{Here the name \textquotedblleft{MIMO}\textquotedblright\ refers to the hardware distortion model whose inputs and outputs are both related to the transmitter side. The inputs are the precoded signals and the outputs are the distorted signals to be transmitted.}. In Fig.~\ref{figdirty}, the backward crosstalk is modeled by the parameters $\kappa_1$ and $\kappa_2$. As mentioned above the backward crosstalk modeled in this way has two origins: crosstalk between in and output transmission lines, and the combination of output crosstalk and impedance mismatch. We have omitted the crosstalk from input to output. It is physically  present due to reciprocity, but the contribution is small since it is not affected by the amplifier's gain.  We consider a symbol-sampled model where the outputs of the transmitter are denoted as $y_1,y_2 \in \mathbb{C}$. The inputs of the transmitter are the OFDM modulated communication signals $x_1,x_2 \in \mathbb{C}$  \cite{PHDR1}, which are modeled as Gaussian distributed. All input signals have the same center frequency.

Let $\bfx= (x_1 \; x_2)^T\in \mathbb{C}^{2\times 1}$ denote the inputs in vector form. When a MIMO transmitter is used for coherent beamforming, the inputs are correlated. To provide a general description, we therefore assume that $\bfx \sim \mathcal{N}_{\mathbb{C}}({\bf 0},{\bf C_x})$, where ${\bf C_x}=\E[\bfx\bfx^H]$ denotes the covariance matrix. A general representation of ${\bf C_x}$ is
\begin{equation} \label{Cx}
{\bf C_x}=\Px\left(
\begin{array}{cc}
1 & \beta\xi  \\
\beta\xi^* & \beta^2    \end{array} \right),
\end{equation}
where the $\Px = \E [ |x_1|^2 ]$ is the power of the first input and it is taken as the reference power in the following parts of this paper. Moreover, $\beta> 0$ is the square root of the ratio between the second signal's power and the first signal's power: $\beta^2 =  \E [ |x_2|^2 ]/ \E [ |x_1|^2 ]$. The correlation coefficient of $x_1$ and $x_2$ is denoted by $\xi \in \mathbb{C}$ and satisfies $|\xi|\leq 1$.     

As shown in Fig.~\ref{figdirty}, the accessible  transmitter output $\bfy= (y_1 \; y_2)^T\in \mathbb{C}^{2\times 1}$ is described by
\begin{equation}\label{y}
  \bfy = \bfr + \bfw,
\end{equation}
where $\bfr = (r_1 \; r_2)^T\in \mathbb{C}^{2\times 1}$ models the output from the power amplifiers and 
$\bfw= (w_1 \; w_2)^T \sim \mathcal{N}_{\mathbb{C}}({\bf 0},\sigma_w^2 \, \bfI)$ models the independent thermal noise that has variance $\sigma_w^2$.

\subsection{Bussgang Description of the Power Amplifier Output}

The input to the $\ell$th power amplifier non-linearity is the internal amplified signal $u_{\ell} \in \mathbb{C}$ and the output is $r_{\ell} \in \mathbb{C}$ for $\ell=1,2$. The input is an internal signal that is not equal to the amplified transmitter input $\gamma_{\ell}x_{\ell}$ when there is backward crosstalk. This is due to the feedback connection between antenna branches as illustrated in Fig.~\ref{figdirty}, i.e., the signal $u_{\ell}$ represents the amplified version of the actual transmitter input, $x_{\ell}$, plus the crosstalk signal that is coupled from the other antenna branch. We will model the backward crosstalk mathematically in Section~\ref{subsec:model-crosstalk}.

The $\ell$th power amplifier will ideally provide an amplification gain of $\gamma_{\ell}>0$, but has a nonlinear behavior determined by the compression parameter $\rho_{\ell} \leq 0$ and the function $f_{\ell}(\cdot)$.
We assume that the power amplifiers are subject to third-order nonlinear distortion that compresses strong input signals, which implies $f_{\ell}(u_{\ell}) = u_{\ell} \, |u_{\ell}|^2$.
 Hence, if $u_{\ell}$ is the input to the $\ell$th amplifier, then the output is
\begin{equation}
r_{\ell} = u_{\ell} + \rho_{\ell}\,f_{\ell}(u_{\ell}) = u_{\ell} + \rho_{\ell} u_{\ell} \, |u_{\ell}|^2, \quad \ell=1,2.
\end{equation}
The compression parameter values $\rho_1,\rho_2$ are typically similar but nonidentical for the two branches.
The vector signal $\bfr$ in (\ref{y}) can then be expressed as
\begin{equation}\label{r}
  \bfr =   \bfu +  \underbrace{  \left(
  \begin{array}{cc}
  \rho_1 & 0  \\
  0 & \rho_2    \end{array} \right) }_{ \triangleq \displaystyle \bfG }
  \underbrace{ \left( \begin{array}{cc}
  u_1 \, |u_1|^2  \\
  u_2 \, |u_2|^2
    \end{array} \right) }_{\triangleq \displaystyle \bff(\bfu)} ,
\end{equation}
where $\bfu = (u_1 \; u_2)^T$ is the input to the nonlinearity. Note that in (\ref{r}), the power amplifier output $\bfr$ is described as a function of the internal signal $\bfu$, where $\bfu$ is a function of the transmitter gains, transmitter input $\bfx$ and the backward crosstalk via the power amplifier output $\bfr$.

By Bussgang decomposition theory, the nonlinear transformation of the input $\bfu$ via $\bff(\bfu) $ in (\ref{r}) can be equivalently described as
\begin{equation} 
\bfr = \bfA \,  \bfu + \bfv  \label{r_Au},
\end{equation}
where $\bfA$ is the constant Bussgang matrix and $\bfv$ is a zero-mean distortion term that is uncorrelated to the input $\bfu$ \cite{PHDR1}. The observation bandwidth of $\bfr$ must be wide enough to comprise all spectral regrowth due to nonlinearities \cite{Ronnow2019}.
The Bussgang matrix $\bfA$ depends on both the properties of the input $\bfu$ exciting the nonlinearity and the nonlinearity $\bff(\bfu)$ itself. As shown in \cite{Emilny}, it can be computed as $\bfA = \E[\bfr \, \bfu^H]
( \E[ \bfu  \, \bfu^H ])^{-1}$ and by substituting \eqref{r_Au} into this expression we obtain
\begin{equation}
 \bfA  =
 \bfI +  \bfG \, \bfUb\, \bfU^{-1}
    \label{eqa},
\end{equation}
where $\bfU \triangleq \E[\bfu \, \bfu^H]$ denotes the covariance matrix of $\bfu$ and  $\bfUb \triangleq \E[ \bff(\bfu)  \, \bfu^H ]$ is a fourth-order moment matrix.
Hence, the Bussgang matrix $\bfA$ in (\ref{eqa}) depends on the transmitter model via $\bfG$ and the second- and fourth-order moments of the internal signal $\bfu$, which are studied below. In addition, the properties of  the power amplifier output $\bfr$ in (\ref{r_Au}) depends also on the nonlinear distortion noise $\bfv$ which is determined later.

\subsection{Modeling the Backward Crosstalk}
\label{subsec:model-crosstalk}

We will now determine the power amplifier input $\bfu$ for the model in Fig.~\ref{figdirty}, where there is backward crosstalk between the transmission lines on the circuit board. This phenomenon is modeled by a feedback network, where
\begin{equation}\label{u}
  \bfu = \underbrace{ \left(
  \begin{array}{cc}
  \gamma_1 & 0   \\
  0 & \gamma_2   \end{array} \right) }_{\triangleq \displaystyle \bfL } \bfx +  \underbrace{ \left(
  \begin{array}{cc}
  0 &   \gamma_1 \, \kappa_2   \\
   \gamma_2 \, \kappa_1  &0   \end{array} \right) }_{\triangleq \displaystyle \bfK }
  \, \bfr .
\end{equation}
Inserting (\ref{r_Au}) into (\ref{u}) yields
\begin{equation}
\bfu = \bfL \, \bfx + \bfK ( \bfA \, \bfu +  \, \bfv).
\end{equation}
By solving for $\bfu$, the signal that excite the non-linearities is obtained as
\begin{equation}\label{ub}
   \bfu =   (\bfI -  \bfK \, \bfA)^{-1} \, ( \bfL \, \bfx  + \bfK
  \, \bfv ).
  \end{equation}
 This signal depends on the transmitter input $\bfx$, the non-linear distortion noise $\bfv$, and on the parameters of the considered transmitter model.
  We study these relationships in further detail in the following example.

  \subsubsection{Linear Model and Transmitter Actual Gain}
To focus on the impact of backward crosstalk, we will now exemplify a linear transmitter with $\bfG=\bfnoll$, symmetric amplification $\gamma=\gamma_1=\gamma_2$, and a common backward crosstalk coefficient $ \delta = \gamma \,  \kappa$ for which $\kappa=\kappa_1=\kappa_2 \in \mathbb{R}$ and $|\delta| < 1$.
  Then, according to (\ref{r}), $\bfr = \bfu$, and using (\ref{ub}) the power amplifier output becomes
  \begin{IEEEeqnarray}{rCl}\label{rb}
   \bfr &=& \frac{1}{1 - \delta^2}
 \left( \! \!  \begin{array}{cc} 1 & \delta \\
   \delta & 1 \end{array} \! \!  \right)  \left(
  \! \!  \begin{array}{cc}
  \gamma & 0   \\
  0 & \gamma   \end{array} \! \!  \right) \, \bfx
  \nonumber \\ &=& \frac{1}{1 - \delta^2}
  \left( \! \!
  \begin{array}{cc}
  \gamma & \delta \, \gamma   \\
  \delta \, \gamma & \gamma   \end{array} \! \!  \right) \, \bfx.  
  \end{IEEEeqnarray}
  Hence, the covariance matrix of $\bfr$ is
  \begin{equation}
  \E [\bfr \, \bfr^H] = \frac{1}{(1 - \delta^2)^2}  \left( \! \!
  \begin{array}{cc}
  \gamma & \delta \, \gamma   \\
  \delta \, \gamma & \gamma   \end{array} \! \!  \right) {\bf C_x}   \left( \! \!
  \begin{array}{cc}
  \gamma & \delta \, \gamma   \\
  \delta \, \gamma & \gamma   \end{array} \! \!  \right). \label{eq:r-covariance}
  \end{equation}
  In appropriately designed radio frequency transmitters, clearly the backward crosstalk yields small errors,
 that is $|\delta|\ll 1$. As $\delta\to 0$, $ \E [\bfr \, \bfr^H] \to \gamma^2 {\bf C_x}$ and the transmitter only provides the amplification gain of $\gamma$. However, it is the case when $|\delta|$ is small but non-zero that is of practical interest.
  
  If the inputs are independent and symmetrically distributed, represented by ${\bf C_x}= \Px \vect{I}$, then for $|\delta|>0$ we notice that \eqref{eq:r-covariance} simplifies to
    \begin{equation}
 \frac{\Px}{(1 - \delta^2)^2}  \left( \! \!
  \begin{array}{cc}
  \gamma^2(1+\delta^2) & 2\delta \, \gamma^2   \\
2\delta \, \gamma^2 & \gamma^2(1+\delta^2)   \end{array} \! \!  \right). \label{eq:r-covariance2}
  \end{equation}
  The non-zero off-diagonal elements in \eqref{eq:r-covariance2} show that the backward crosstalk makes the outputs correlated even when the inputs are uncorrelated. From \eqref{eq:r-covariance2}, we also notice that the output power $\sigma_{r\ell}^2 = \E[|r_\ell|^2]$ of the $\ell$th output is
 \begin{IEEEeqnarray}{rCl}
  \sigma_{r\ell}^2 =  \frac{\gamma^2 \, (1 + \delta^2)}{(1 - \delta^2)^2} \, \Px
   \DEF
   \bar \gamma^2 \, \Px, \label{bargamma}
   \end{IEEEeqnarray}
where $\bar \gamma = \gamma \sqrt{1 + \delta^2}/(1 - \delta^2)$ is the actual amplification gain of the transmitter. We have $\bar \gamma \simeq \gamma$ for $|\delta| \ll 1$, where $\simeq$ denotes an approximate expression where only the dominant terms are retained. Nevertheless, we want to further understand the small errors that also occur in the regime of $|\delta| \ll 1$.

\subsection{Small-Error Analysis}

In (\ref{ub}), we have derived that the internal signal $\bfu$ is a weighted sum of the input $\bfx$ and the non-linear distortion noise $\bfv$, which is by construction uncorrelated to $\bf u$. Although $\bfx$ was assumed to be Gaussian, the distortion noise $\bfv$ is non-Gaussian and also not independent of $\bfx$, which makes an exact analysis cumbersome. In a small-error analysis where the distortion noise is negligible, 
 the signal $\bfu$ in (\ref{ub}) can be approximated  as
\begin{equation}\label{u_small_error}
   \bfu \simeq  (\bfI -  \bfK \, \bfA)^{-1} \, \bfL    \, \bfx.
\end{equation}
When the matrix representing the compression of power amplifiers, ${\bf G}$ is zero (i.e., the power amplifiers are ideal), the above approximation becomes exact. Note that this is a meaningful assumption when the power amplifiers operate close to the linear operation, which is possible with sufficient input power back-off. Furthermore, for a transmitter working close to its linear operation it holds that  $\bfK \bfA \simeq \bfK$ and $|\gamma_1\gamma_2\kappa_1\kappa_2|\ll1$, so that
\begin{equation}\label{u_small_errorb}
   \bfu \simeq
  \underbrace{ \left(
  \begin{array}{cc}
  \gamma_1 & \gamma_1 \, \kappa_2 \, \gamma_2\\
  \gamma_2 \, \kappa_1 \, \gamma_1 & \gamma_2 \end{array} \right)}_{\DEF \displaystyle \bfQ} \, \bfx.
\end{equation}
From now on, we will utilize this approximate description, for which we can compute the Bussgang matrix $\vect{A}$ in closed form. The linear relation in \eqref{u_small_errorb} is based on two approximations: 1) the power amplifier operates close to its linear region, and 2) the power of the backward crosstalk is much smaller than the actual desired signal. These are all reasonable since the hardware is pre-calibrated to limit the distortion and only the small residual impairments remain to be modeled. In this paper, we analyze the system mathematically under these approximations and we then provide numerical simulations in Section~\ref{numerical} to demonstrate the accuracy.

From (\ref{u_small_errorb}), we first notice that the internal signal $\bfu$ is now Gaussian distributed with covariance matrix $\bfU \DEF \E[\bfu \, \bfu^H]$ given by
\begin{IEEEeqnarray}{rCl}\label{U}
\bfU
&\DEF&  \left(
  \begin{array}{cc}
  u_{11} & u_{12}   \\
  u_{12}^* & u_{22}   \end{array} \right) = \bfQ \, {\bf C_x} \, \bfQ^H,
\end{IEEEeqnarray}
where the elements of $\bfU$ are denoted as
\begin{IEEEeqnarray}{rCl}
&	u_{\ell\ell} & \DEF \E[u_\ell \, u_\ell^*]=t_{\ell\ell}P_x, \quad \ell=1,2,  \nonumber \\
&	u_{12} & \DEF \E[u_1 \, u_2^*]=t_{12}P_x , 
\end{IEEEeqnarray}
where $t_{11}$, $t_{12}$, $t_{22}$ can be computed as
\begin{IEEEeqnarray}{rCl}
	t_{11} &\DEF& \gamma_1^2 +2\gamma_1^2\gamma_2 \beta  \Re\{ \kappa_2^{*}\xi\}+\gamma_1^2 \gamma_2^2 |  \kappa_2 |^2\beta^2 \label{t11n} ,  \\
	t_{12} &\DEF& \gamma_1 \gamma_2 \left(
	\gamma_1 \kappa_1^*  +\beta\xi+\gamma_1 \gamma_2  \kappa_1^{*} \kappa_2\beta\xi^*+ \gamma_2  \kappa_2 \beta^2
	\right) , \label{t12} \\
	t_{22} &\DEF& \gamma_2^2\beta^2 +2\gamma_1\gamma_2^2\beta\Re\{ \kappa_1\xi\}+ \gamma_1^2 \gamma_2^2 |  \kappa_1 |^2 \label{t22n}.
\end{IEEEeqnarray}

When the inputs to the non-linearity is Gaussian distributed, it follows that the Bussgang matrix $\vect{A}$ is diagonal \cite[Sec.~II.B]{Emilny} and given by
\begin{align}
	\bfA&=\left( \!\!
	\begin{array}{cc}
	a_1 & 0    \\
	0 &a_2  \end{array} \!\!\right)=\left( \!\!
	\begin{array}{cc}
		\frac{\E[ r_1 u_1^* ]}{\E[ |u_1|^2 ]} & 0    \\
		0 & \frac{\E[ r_2 u_2^* ]}{\E[ |u_2|^2 ]}  \end{array} \!\!\right) \nonumber \\ 
		&=\left( \!\!
		\begin{array}{cc}
		1+2\rho_1u_{11} & 0    \\
		0 & 1+2\rho_2u_{22}   \end{array} \!\! \right) \label{A}.
\end{align}
This result alternatively can be obtained by the first part of Appendix~\ref{V_matrix}. Since $\rho_1, \rho_2$ are negative parameters, the diagonal elements of $\vect{A}$ will be smaller than one. For practical parameter values it holds that $0<|1+2\rho_{\ell} u_{\ell \ell}|<1$, for $\ell=1,2$, thus when substituting \eqref{A} into \eqref{r_Au}, the output of the nonlinearity is the input with reduced gain plus the nonlinear distortion noise. We call $1+2\rho_{\ell} u_{\ell \ell}$ the Bussgang attenuation.

\subsection{Transmitter Output Error}

We will now utilize the Bussgang decomposition in (\ref{r_Au}), the closed-form results in the small-error regime,  and the  obtained properties of the internal signal $\bfu$ to derive properties of the transmitter output error. These properties are key to analyze the end performance of the transmitter.

The transmitter output $\bfy$ in (\ref{y})  is now given as functions of the transmitter input $\bfx$, nonlinear distortion noise  $\bfv$, and  transmitter thermal noise $\bfw$ by
\begin{equation}\label{y2}
\bfy = \bfA \, \bfQ \, \bfx + \bfv + \bfw,
 \end{equation}
 where (\ref{r_Au}) and (\ref{u_small_errorb}) were used to form an expression of the input $\bfx$ and nonlinear distortion noise $\bfv$.
 In (\ref{y2}), $\bfQ$ is given in (\ref{u_small_errorb}) and the matrix $\bfA$ in (\ref{A}). The ideal output $\bfy_{\mbox{\scriptsize o}}$ of the transmitter is given by the purely amplified inputs, that is
\begin{equation}
\bfy_{\mbox{\scriptsize o}} = \left(
  \begin{array}{c}
   y_{\mbox{\scriptsize o} 1}  \\
   y_{\mbox{\scriptsize o} 2} \end{array} \right) \triangleq \bfL \, \bfx, \label{y0}
\end{equation}
 where the gain matrix $\bfL$ is defined in (\ref{u}). From (\ref{y2}) and (\ref{y0}), the error signal $\bfe$ determining the full properties of the $2 \times 2$ MIMO transmitter reads
\begin{equation} \label{error}
  \bfe \DEF \bfy - \bfy_{\mbox{\scriptsize o}} =
  \underbrace{( \bfA \, \bfQ - \bfL) \,  \bfx}_{\triangleq \displaystyle \bfxt} + \bfv + \bfw.
\end{equation}
The introduced signal $\bfxt$ captures (the linear part of) the error due to backward crosstalk and Bussgang attenuation, $\bfv$ is the nonlinear distortion noise, and $\bfw$ is the thermal noise.  The model (\ref{error}) of the transmitter output error $\bfe$ forms the basis for the performance analysis in the following sections.

\section{Performance Analysis Based on the NMSE \label{nmse}}

One way to measure the transmitter performance is to measure the NMSE between the ideal output $\bfy_{\mbox{\scriptsize o}}$ in \eqref{y0} and the true output in \eqref{y2}. This is the normalized power of the error vector in \eqref{error}.
The three error terms defining the total error in (\ref{error}) are jointly uncorrelated. Accordingly,  the error covariance $\bfE = \E[ \bfe \, \bfe^H]$ can be written as

\begin{equation} \label{E}
 \bfE  =  \bfXt + \bfV  + \sigma_w^2 \, \bfI,
\end{equation}
where the covariance matrices $\bfXt = \E [\bfxt \, \bfxt^H] $ and $\bfV = \E [\bfv \, \bfv^H] $ are to be determined, while $\sigma_w^2$ still denotes the variance of the thermal noise.

\subsection{Properties of the Linear Error Matrix $\bfXt$  }

From (\ref{error}) it follows that  the covariance matrix $\bfXt$ in (\ref{E}) reads
\begin{equation} \label{Xtilde}
\bfXt = (\bfA \, \bfQ - \bfL) {\bf C_x} (\bfA \, \bfQ - \bfL)^H,
\end{equation}
where $\E[\bfx \, \bfx^H]={\bf C_x}$ was used. It follows from a straightforward calculation using $\bfA$ in (\ref{A}), $\bfQ$ in (\ref{u_small_errorb}), and $\bfL$ in (\ref{u}) that
\begin{equation} \label{int1}
 \bfA \, \bfQ - \bfL =
\left(\!\!
  \begin{array}{cc}
   2 \gamma_1 \, \rho_1 \, u_{11} &
   \gamma_1 \, \kappa_2 \, \gamma_2(1+2\rho_1u_{11})\\
   \gamma_2 \, \kappa_1 \, \gamma_1(1+2\rho_2u_{22}) &
   2 \gamma_2 \, \rho_2  \, u_{22} \end{array} \!\!\right),
\end{equation}
where the result in (\ref{int1}) is expressed in terms of the intermediate diagonal elements of $\bfU$ in (\ref{U}), and the transmitter parameters. We can also express $\bfA\bfQ-\bfL$ as
\begin{equation} \label{AQ_L}
\bfA\bfQ-\bfL=2\bfG\bfB\bfQ+\bfK\bfL,
\end{equation}
where $\bfK$ is defined in (\ref{u}) and $\bfB$ is defined as
\begin{equation} \label{B}
\bfB \DEF \left(
\begin{array}{cc}
u_{11} &
0\\
0 &
u_{22} \end{array}  \right).
\end{equation}
Using (\ref{U}) and (\ref{AQ_L}), the covariance matrix $\bfXt$ in (\ref{Xtilde}) reads 
\begin{IEEEeqnarray}{rCl}
\bfXt&=&4\bfG\bfB\bfU\bfB\bfG+\bfK\bfL{\bf C_x}\bfL\bfK^H \nonumber \\ &&+2\bfK\bfL{\bf C_x}{\bf Q}^H\bfB\bfG+2\bfG\bfB\bfQ{\bf C_x}\bfL\bfK^H.
\end{IEEEeqnarray}
If the elements of $\bfXt$ are denoted as
\begin{equation}
 \bfXt  \DEF  \left(
  \begin{array}{cc}
  \tilde x_{11} & \tilde x_{12}   \\
  \tilde x_{12}^* & \tilde x_{22}   \end{array} \right),
\end{equation}
then a straightforward calculation provides 
\begin{IEEEeqnarray}{rCl}
\label{xb11n} \tilde x_{11} &=&
4   \rho_1^2 \, t_{11}^3 \, \Px^3+4\gamma_1^2\gamma_2t_{11}\rho_1\left(\gamma_2\beta^2\left|\kappa_2\right|^2+\beta\Re\left\{\kappa_2\xi^*\right\}\right)\Px^2 \nonumber \\
&&+\beta^2\gamma_1^2\gamma_2^2|\kappa_2|^2\Px,
\\ \label{xb12n} \tilde x_{12} &=&  4   \rho_1 \, \rho_2 \, t_{12} \, t_{11} \, t_{22} \, \Px^3+2\gamma_1^2\gamma_2t_{11}\rho_1\kappa_1^*\left(1+\gamma_2\beta\xi^*\kappa_2\right)\Px^2\nonumber \\
&&+2\gamma_1\gamma_2^2t_{22}\rho_2\kappa_2\left(\beta^2+\gamma_1\beta\xi^*\kappa_1^*\right)\Px^2 \nonumber \\
&&+\beta\xi^*\gamma_1^2\gamma_2^2\kappa_1^*\kappa_2\Px , \\
\label{xb22n} \tilde x_{22} &=& 4   \rho_2^2 \, t_{22}^3 \, \Px^3+4\gamma_1\gamma_2^2t_{22}\rho_2\left(\gamma_1\left|\kappa_1\right|^2+\beta\Re\left\{\kappa_1\xi\right\}\right)\Px^2 \nonumber \\ &&+\gamma_1^2\gamma_2^2|\kappa_1|^2\Px.
\end{IEEEeqnarray}
In the above equations, the results are compactly expressed in terms of the input signal and transmitter hardware parameters. Note that all the terms in (\ref{xb11n})-(\ref{xb22n}) are third-order polynomials of the reference input power $\Px$. The result above will be used to derive the error covariance in (\ref{E}), once the elements of $\bfV$ have been determined.

\subsection{Properties of the Nonlinear Error Matrix $\bfV$}

The covariance matrix of the distortion noise $\bfv$ is known to be \cite{PHDR1}
\begin{equation} \label{V}
\bfV =
\bfG \, ( \bfUbb -  \bfUb \, \bfU^{-1} \, \bfUb^H  ) \, \bfG^H,
\end{equation}
where $\bfG$ is given in (\ref{r}) and $\bfUbb = \E[ \bff(\bfu) \, \bff(\bfu)^H ]$ is the matrix of higher-order moments. It is shown in Appendix~\ref{V_matrix} that (\ref{V}) can be reduced to
\begin{equation} \label{V2}
\bfV =
2 \, \bfG \, \bfC\, \bfG^H,
\end{equation}
where the matrix $\bfC$ can be compactly written using the properties of the internal signal $\bfu$:
\begin{IEEEeqnarray}{rCl}\label{C}
\bfC = \left(   \begin{array}{cc}
u_{11}^3 &  u_{12} \, |u_{12}|^2 \\ u_{12}^* \, |u_{12}|^2 &   u_{22}^3
\end{array}   \right) .
\end{IEEEeqnarray}
With (\ref{V2}) and (\ref{C}) as starting point,  a straightforward calculation  reveals that the covariance matrix $\bfV$ in (\ref{V})
reads
\begin{equation} \label{V3}
\bfV \DEF  \left(
  \begin{array}{cc}
  v_{11} &v_{12}   \\
  v_{12}^* & v_{22}   \end{array} \right),
\end{equation}
where
\begin{IEEEeqnarray}{rCl}
 v_{\ell\ell} &=&  2 \rho_\ell^2 \, u_{\ell\ell}^3=2\rho_\ell^2t_{\ell\ell}^3\Px^3, \quad  \ell=1,2,  \label{v11n} \\
v_{12} &=&  2 \rho_1  \, \rho_2 \,  u_{12} \, |u_{12}|^2=2\rho_1\rho_2t_{12}|t_{12}|^2\Px^3. \label{v12n} 
\end{IEEEeqnarray}

\subsection{Closed-form Expressions for the NMSE}

By denoting the diagonal elements of  the error covariance $\bfE$ in (\ref{E}) as  $e_{1 1}$, $e_{2 2}$, the figure-of-merit NMSE for the first and second branch is given as
\begin{IEEEeqnarray}{rCl}
\NMSE_1 &\DEF& \frac{e_{1 1}}{\mathbb{E}\{ | y_{\mbox{\scriptsize o} 1} |^2\}} = 
\frac{e_{1 1}}{\gamma_1^2 \, \Px},  \label{NMSE1} \\
\NMSE_2 &\DEF&  \frac{e_{2 2}}{\mathbb{E}\{ | y_{\mbox{\scriptsize o} 2} |^2\}} = 
\frac{e_{2 2}}{\gamma_2^2 \, \beta^2 \, \Px}.  \label{NMSE2} 
\end{IEEEeqnarray}
By utilizing the expressions derived in \eqref{xb11n}, \eqref{xb22n}, \eqref{v11n}, the diagonal elements $e_{11}$ and $e_{22}$ are obtained as
\begin{IEEEeqnarray}{rCl} 
e_{11}  &=&
 \tilde x_{11} + v_{11}     + \sigma_w^2 \nonumber \\
 &=&6   \rho_1^2 \, t_{11}^3 \, \Px^3+4\gamma_1^2\gamma_2t_{11}\rho_1\left(\gamma_2\beta^2\left|\kappa_2\right|^2+\beta\Re\left\{\kappa_2\xi^*\right\}\right)\Px^2\nonumber \\
 &&+\beta^2\gamma_1^2\gamma_2^2|\kappa_2|^2\Px+\sigma_w^2, \label{e11} \\
 e_{22}  &=&
 \tilde x_{22} + v_{22}     + \sigma_w^2 \nonumber \\
 &=&6   \rho_2^2 \, t_{22}^3 \, \Px^3+4\gamma_1\gamma_2^2t_{22}\rho_2\left(\gamma_1\left|\kappa_1\right|^2+\beta\Re\left\{\kappa_1\xi\right\}\right)\Px^2\nonumber\\
 &&+\gamma_1^2\gamma_2^2|\kappa_1|^2\Px+\sigma_w^2. \label{e22} 
\end{IEEEeqnarray}
With the variances $e_{11}$ and $e_{22}$ of the first and second branch errors in (\ref{e11})-(\ref{e22}), $\NMSE_1$ and $\NMSE_2$ becomes
\begin{IEEEeqnarray}{rCl} 
 \NMSE_1 &=&  \frac{6\rho_1^2  \, t_{11}^3}{\gamma_1^2} \, \Px^2 \nonumber\\&&+4\gamma_2t_{11}\rho_1\left(\gamma_2\beta^2\left|\kappa_2\right|^2+\beta\Re\left\{\kappa_2\xi^*\right\}\right)\Px\nonumber\\&&+\beta^2\gamma_2^2|\kappa_2|^2+\frac{\sigma_w^2}{\gamma_1^2 \, \Px}, \label{NMSE1b} \\
 \NMSE_2 &=&     \frac{6\rho_2^2  \, t_{22}^3}{\gamma_2^2\beta^2} \, \Px^2 \nonumber\\&&+\frac{4\gamma_1t_{22}\rho_2}{\beta^2}\left(\gamma_1\left|\kappa_1\right|^2+\beta\Re\left\{\kappa_1\xi\right\}\right)\Px \nonumber\\&&+\gamma_1^2\frac{|\kappa_1|^2}{\beta^2}+\frac{\sigma_w^2}{\gamma_2^2 \, \beta^2 \Px}. \label{NMSE2b}
\end{IEEEeqnarray}

\begin{lemma} \label{lemma:convex-functions}
The NMSEs for the two antenna branches, $\NMSE_1$ and $\NMSE_2$, are convex functions of the reference input power $\Px$ for $\Px \geq 0$. 
\end{lemma}
\begin{IEEEproof}
This can easily be proved by taking the second derivative of $\NMSE_1(\Px)$ and $\NMSE_2(\Px)$ with respect to $\Px$ and show that they are positive. Direct differentiation yields
\begin{align}
& \NMSE_1''(\Px)= \frac{12\rho_1^2  \, t_{11}^3}{\gamma_1^2}+\frac{2\sigma_w^2}{\gamma_1^2 \, \Px^3},\nonumber\\& \NMSE_2''(\Px)=\frac{12\rho_2^2  \, t_{22}^3}{\gamma_2^2\beta^2}+\frac{2\sigma_w^2}{\gamma_2^2 \, \beta^2 \Px^3}. \label{NMSE2_der}
\end{align}
Both are positive for $\Px\geq0$ which is the range of interest.
\end{IEEEproof}

The closed-form NMSE expressions in \eqref{NMSE1b} and \eqref{NMSE2b} provide insights into the hardware behavior. Let us focus on $\NMSE_1$ since the other branch has identical characteristics, except for the different notation. If we assume that the strength of the backward crosstalk signal is very small compared to the main signal, we have $\gamma_2 \left|\kappa_2\right| \ll 1$ and $t_{11}$ in \eqref{t11n} can be approximated as $t_{11} \simeq \gamma_1^2$ and furthermore $\NMSE_1$ can be approximated as
 \begin{align}
 \NMSE_1 \simeq  6\rho_1^2\gamma_1^4  \, \Px^2+4\rho_1\beta\gamma_2\Re\left\{\kappa_2\xi^*\right\}\gamma_1^2\Px +\frac{\sigma_w^2}{\gamma_1^2 \, \Px}.\label{NMSE1-2}
\end{align}
We note that it is a convex function of the ideal amplified signal power $\mathbb{E}\{ | y_{\mbox{\scriptsize o} 1} |^2\} =\gamma_1^2\Px$. This means that when the desired amplifier gain of the first transmitter branch, $\gamma_1$ increases, the input reference power $\Px$ should be decreased at the same ratio with $\gamma_1^2$ in order to keep the NMSE the same. Moreover, the first term in \eqref{NMSE1-2} is a monotonically increasing function of $\Px$ and as the compression ratio of the first branch's power amplifier, $|\rho_1|$, increases, it increases with the square of it. Note that this term dominates the NMSE when $\gamma_1^2\Px$ grows. The sign of the second term is dependent on $\Re\left\{\kappa_2\xi^*\right\}$. Since $\rho_1\leq0$, when the crosstalk parameter $\kappa_2$ and the correlation coefficient $\xi$ are phase-aligned, this term reduces the NMSE to some extent. In this case, the power amplifier non-linearity corresponding to the first term is the main source for the distortion. When $\Re\left\{\kappa_2\xi^*\right\}<0$, the sum of the first two terms monotonically increases with $\Px$ and we see a combination of power amplifer and backward crosstalk distortion. In this case, the last term regularizes the NMSE since it decreases with $\Px$ and goes to infinity as $\Px \to 0$, hence the optimal input reference power is clearly non-zero. This term dominates the NMSE expression when $\Px$ is small compared to the thermal noise variance.

\subsection{Power Back-off for Minimizing Maximum NMSE}

As discussed above, the NMSE of an antenna branch is minimized at a non-zero value of $\Px$. For a single-antenna transceiver, there is only one NMSE and therefore it is desirable to find the average input power that minimizes its NMSE.
We can study that special case by setting $\beta =0$, we obtain
$ \NMSE_1 =  6\rho_1^2  \, \gamma_1^4 \, \Px^2 +\frac{\sigma_w^2}{\gamma_1^2 \, \Px}$ from \eqref{NMSE1b}. It is then straightforward to show that it is minimized by
\begin{equation}
\Px = \frac{1}{\gamma_1^2} \sqrt[3]{ \frac{\sigma_w^2}{12 \rho_1^2 }},
\end{equation}
which depends on the compression parameter as $1/|\rho_1|^{2/3}$.

Since we have a 2$\times$2 MIMO transmitter structure with two different NMSE expressions, given in \eqref{NMSE1b} and \eqref{NMSE2b}, there is generally not one value of $\Px$ that jointly minimizes both NMSEs. Hence, we take a min-max fair optimization approach that minimizes the maximum of $\NMSE_1$ and $\NMSE_2$. The optimization problem for this aim can be cast as
\begin{align}
& \underset{\Px,\epsilon}{\text{minimize}} \ \ \ \epsilon \label{objective} \\
& \text{subject to} \ \ \ \NMSE_1(\Px)\leq \epsilon \nonumber \\
& \hspace{1.8cm} \NMSE_2(\Px)\leq \epsilon \nonumber,
\end{align}  
where $\epsilon$ represents the maximum NMSE value of the two branches.
This is a convex optimization problem since the cost function is linear and the NMSEs are convex functions of $\Px$, as proved in Lemma~\ref{lemma:convex-functions}. Hence, the problem can be solved numerically using standard convex optimization solvers. However, the following theorem presents the optimal closed-form input reference power, $\Px$, for the problem (\ref{objective}).

\begin{thm} \label{thm1}
The optimal input reference signal power, $\Px$ for the problem (\ref{objective}) is given by
\begin{align}
& \Px^{\text{NMSE-opt}}=\text{arg} \min_{\Px\in \mathcal{S}} \ \ \max\bigg\{\NMSE_1(\Px), \NMSE_2(\Px)\bigg\},
\end{align}
where $\mathcal{S}=\Big\{\Px^{(1)},\Px^{(2)},\Px^{(3)}\Big\}$ and the elements of the set $\mathcal{S}$ are given as follows:

1) $\Px^{(1)}$ is the unique positive root of the third-order polynomial
   \vspace{-2mm}
\begin{align}
&12\rho_1^2  \, t_{11}^3 \, \Px^3 \nonumber\\&+ 4\gamma_1^2\gamma_2t_{11}\rho_1\left(\gamma_2\beta^2\left|\kappa_2\right|^2+\beta\Re\left\{\kappa_2\xi^*\right\}\right)\Px^2-\sigma_w^2. \label{NMSE1_polyn}
\end{align}

2) $\Px^{(2)}$ is the unique positive root of the third-order polynomial
\begin{align}
& 12\rho_2^2  \, t_{22}^3 \, \Px^3 + 4\gamma_1\gamma_2^2t_{22}\rho_2\left(\gamma_1\left|\kappa_1\right|^2+\beta\Re\left\{\kappa_1\xi\right\}\right)\Px^2-\sigma_w^2. \label{NMSE2_polyn} 
\end{align}

3) If it exists,  $\Px^{(3)}$  is the positive root of the third-order polynomial 
\begin{align}
&\frac{6\left(\rho_1^2  \, t_{11}^3 \, \gamma_2^2\beta^2-\rho_2^2  \, t_{22}^3\gamma_1^2\right)}{\gamma_1^2\gamma_2^2\beta^2} \, \Px^3\nonumber\\&+4\gamma_2t_{11}\rho_1\left(\gamma_2\beta^2\left|\kappa_2\right|^2+\beta\Re\left\{\kappa_2\xi^*\right\}\right)\Px^2\nonumber \\
&-\frac{4\gamma_1t_{22}\rho_2}{\beta^2}\left(\gamma_1\left|\kappa_1\right|^2+\beta\Re\left\{\kappa_1\xi\right\}\right)\Px^2\nonumber\\&+\left(\beta^2\gamma_2^2|\kappa_2|^2-\gamma_1^2\frac{|\kappa_1|^2}{\beta^2}\right)\Px+\frac{\sigma_w^2}{\gamma_1^2}-\frac{\sigma_w^2}{\gamma_2^2 \, \beta^2}, \label{case3}
\end{align}
 which makes $\NMSE_1=\NMSE_2$ the smallest.
 \end{thm}
\begin{IEEEproof}
	The proof is provided in Appendix~\ref{appthe1}.
\end{IEEEproof}

We call the optimal solution provided by Theorem~\ref{thm1} a closed-form solution since no optimization needs to be carried out to obtain it. We only need to compute the NMSEs for the three candidate solutions in $\mathcal{S}$ and pick the one that minimizes the maximum of the NMSEs of the two transmitter branches.
We stress that the roots of third-order polynomials are available in closed form \cite{Hungerford1996a} but we will not give these expressions here since they are lengthy and  offer no additional insights. We will instead analyze the solutions numerically in Section~\ref{numerical}.

\section{Spectral Efficiency of $2 \times 1$ MISO Channel \label{se}}

In this section, we turn the attention to the receiver side by considering the impact that the transmitter distortion (i.e., nonlinearity and backward crosstalk) has on the communication performance, characterized by the SE. More precisely, we consider a $2 \times 1$ multiple-input single-output (MISO) channel\footnote{In this section, different from the previous parts, the name ``MISO'' refers to the $2\times 1$ physical channel between the outputs of the $2\times 2$ dirty MIMO transmitter and the single-antenna receiver.} where the signals sent from the antennas are $y_1$ and $y_2$, which is given in vector form as in \eqref{y2}, while the received signal at the single-antenna receiver is
\begin{align}\label{z}
z={\bf h}^T{\bf y}+n,
\end{align}
where $n\sim \mathcal{N}_{\mathbb{C}}\left(0,\sigma_n^2\right)$ is the independent receiver noise, which might also include interference. The vector ${\bf h}=( h_1 \ h_2 )^T\in \mathbb{C}^{2 \times 1}$ represents the equivalent baseband channel from transmitter to the receiver, where $h_1$ and $h_2$ are the channel coefficients from the first and second transmit antenna, respectively. Since our main aim is to quantify the impact of transmitter distortion, we assume the channel coefficients are deterministic and known, and we have also assumed that the receiver hardware is ideal.
The data is encoded into the transmitted signal ${\bf y}$ by selecting the input signal ${\bf x}$. Since we consider a single-antenna receiver, we consider precoded transmission:
   \vspace{-6mm}
\begin{align}
{\bf x}={\bf c}\bar{x}, \label{input}
\end{align}
with ${\bf c}=( c_1 \ c_2 )^T\in \mathbb{C}^{2\times 1}$ being the fixed precoding  vector and $\bar{x}$ is a scalar data signal. Since the SE is maximized by Gaussian data codebooks \cite{Emilbook}, we assume that $\bar{x} \sim\mathcal{N}_{\mathbb{C}}(0,1)$ and, thus, the input ${\bf x}$ is a zero-mean complex Gaussian vector with covariance matrix 
\begin{align} \label{Cx2}
{\bf C_x}={\bf c}{\bf c}^H=\left(
\begin{array}{cc}
|c_1|^2 & c_1c_2^*  \\
c_1^*c_2 & |c_2|^2    \end{array} \right).
\end{align}
When comparing \eqref{Cx2} with the original model in \eqref{Cx}, we can identify $\Px=|c_1|^2$, $\beta=\frac{|c_2|}{|c_1|}$, $\xi=e^{j\angle{c_1c_2^*}}$. Using \eqref{y2} and \eqref{z}, we can express the received signal $z$ as
\begin{align} \label{z2}
z={\bf h}^T{\bf A}{\bf Q}{\bf c}\bar{x}+{\bf h}^T{\bf v}+{\bf h}^T{\bf w}+n,
\end{align}
where the first term is the desired signal term and the other three terms are noise that are mutually uncorrelated with the desired signal and each other. However, the effective noise, ${\bf h}^T{\bf v}+{\bf h}^T{\bf w}+n$, is not Gaussian distributed, hence the exact channel capacity is hard to obtain. However, we can use a well-known result  \cite[Corollary 1.3]{Emilbook} to obtain the following lower bound on the capacity:
it is well known that the capacity of $2\times 1$ MISO channel is lower bounded by 
\begin{align}\label{Rtilde}
\bar{R}&\DEF \log_2\left(1+\frac{\left|{\bf h}^T{\bf A}{\bf Q}{\bf c}\right|^2}{\mathbb{E}\left\{\left|{\bf h}^T{\bf v}\right|^2\right\}+\mathbb{E}\left\{\left|{\bf h}^T{\bf w}\right|^2\right\}+\sigma_n^2}\right) \nonumber\\&
= \log_2\left(1+\frac{\left|{\bf h}^T{\bf A}{\bf Q}{\bf c}\right|^2}{{\bf h}^T{\bf V}{\bf h}^*+\sigma_w^2{\bf h}^H{\bf h}+\sigma_n^2}\right).
\end{align}
This is called an achievable SE and is measured in bit per channel use. Note that in \eqref{Rtilde}, the matrix ${\bf Q}$, the channel vector ${\bf h}$, and the transmitter and receiver noise powers, $\sigma_w^2$ and $\sigma_n^2$ represent constant system parameters that are independent of the precoding vector ${\bf c}$, and hence ${\bf C_x}$. The expression is valid for any precoding vector ${\bf c}$, but it is desirable to identify the precoding that maximizes the SE.

In a distortion-free system, the SE is maximized by maximum ratio transmission (MRT) \cite{Lo1999a} for which ${\bf c}$ is a scaled version of the conjugate channel ${\bf h}^*$, where the scaling determines the transmit power. It is then desirable to transmit at as high power as possible to maximize the SE. None of these conventional properties hold under the considered transmitter distortion model, thus we will select ${\bf c}$ to maximize $\bar{R}$.

\subsection{Precoder Design for Maximizing Spectral Efficiency}

We will now optimize ${\bf c}$ to maximize $\bar{R}$, which is equivalent to maximizing the signal-to-noise-plus-distortion ratio (SNDR) inside the logarithm in \eqref{Rtilde}:
\begin{equation} \label{sndr}
\mathrm{SNDR} = \frac{\left|{\bf h}^T{\bf A}{\bf Q}{\bf c}\right|^2}{{\bf h}^T{\bf V}{\bf h}^*+\sigma_w^2{\bf h}^H{\bf h}+\sigma_n^2}.
\end{equation}
Due to the backward crosstalk, the input ${\bf u}$ to the power amplifiers has covariance matrix 
\begin{equation}
{\bf U}=\bfQ \, {\bf C_x} \, \bfQ^H = \bfQ {\bf c}{\bf c}^H \bfQ^H,
 \end{equation}
 where the matrix $\bfQ$ is defined in \eqref{u_small_errorb}.
 We therefore define the effective precoding vector ${\bf \tilde{c}}\DEF {\bf Q}{\bf c}$ and note that ${\bf U}$ becomes
 \begin{align}
{\bf U}=\left(\begin{array}{cc} u_{11}  & u_{12} \\ u_{12}^* & u_{22} \end{array}\right)={\bf \tilde{c}}{\bf \tilde{c}}^H=\left(\begin{array}{cc} |\tilde{c}_1|^2  & \tilde{c}_1\tilde{c}_2^* \\ \tilde{c}_1^*\tilde{c}_2 & |\tilde{c}_2|^2 \end{array}\right).\label{Urate}
\end{align}
Since  $\bfQ$ is an invertible matrix, we can without loss of generality maximize  $\mathrm{SNDR}$ with respect to ${\bf \tilde{c}}$ instead.
With this new notation, the matrices ${\bf A}$ and ${\bf V}$ in the numerator and denominator of $\mathrm{SNDR}$ can be expressed as
\begin{align}
&{\bf A}=\left(\begin{array}{cc} 1+2\rho_1|\tilde{c}_1|^2  & 0 \\ 0 & 1+2\rho_2|\tilde{c}_2|^2 \end{array}\right),\nonumber\\& {\bf V}=\left(\begin{array}{cc} 2\rho_1^2|\tilde{c}_1|^6  & 2\rho_1\rho_2|\tilde{c}_1|^2|\tilde{c}_2|^2\tilde{c}_1\tilde{c}_2^* \\ 2\rho_1\rho_2|\tilde{c}_1|^2|\tilde{c}_2|^2\tilde{c}_1^*\tilde{c}_2 &  2\rho_2^2|\tilde{c}_2|^6  \end{array}\right).\label{Arate}
\end{align}
Using (\ref{Arate}), the SE maximization problem can be equivalently expressed as
\begin{align} \label{rate_max}
&\underset{\tilde{c}_1,\ \tilde{c}_2}{\text{maximize}} \ \ \ \frac{2\left| h_1\tilde{c}_1+h_2\tilde{c}_2+\tilde{h}_1|\tilde{c}_1|^2\tilde{c}_1+\tilde{h}_2|\tilde{c}_2|^2\tilde{c}_2 \right|^2}{\left|\tilde{h}_1\left|\tilde{c}_1\right|^2\tilde{c}_1+\tilde{h}_2\left|\tilde{c}_2\right|^2\tilde{c}_2\right|^2+\sigma^2}
\end{align}
where the following constants were defined for ease of notation:
\begin{align}
&\tilde{h}_{\ell} \DEF 2h_{\ell} \rho_{\ell}, \quad  \ell=1,2, \ \ \ \ \  \sigma^2 \DEF 2\sigma_w^2{\bf h}^H{\bf h}+2\sigma_n^2. \label{htilde}
\end{align}
The optimal precoder weights are computed as follows. 

\begin{thm} \label{thm:optimal-precoding}
The precoding vector that maximizes the SE is given by ${\bf c}={\bf Q}^{-1}{\bf \tilde{c}}^{\mathrm{opt}}$, where ${\bf \tilde{c}}^{\mathrm{opt}} = (\tilde{c}_1^{\mathrm{opt}} \; \tilde{c}_2^{\mathrm{opt}})^T$ is computed as
\begin{align} \label{eq:SE-maximization}
\left\{\tilde{c}_1^{\mathrm{opt}}, \tilde{c}_2^{\mathrm{opt}}\right\}=\text{arg} \max_{\left\{\tilde{c}_1,\tilde{c}_2\right\}\in \mathcal{C}} \ \ \bar{R}(\tilde{c}_1,\tilde{c}_2),
\end{align}
where the set $\mathcal{C}$ is given by
\begin{align}
\mathcal{C}=&\Bigg\{     \left\{ 0, \varrho_2^{(\text{1-B-1})} \right\}, \left\{ 0, \varrho_2^{(\text{1-B-2})} \right\},      \left\{\varrho_1^{(\text{1-C-1})},0 \right\} , \nonumber\\& \left\{ \varrho_1^{(\text{1-C-2})},0\right\}, \left\{\varrho_1^{(\text{1-C-1})}e^{j\angle{h_1^*h_2}},\sqrt{\frac{\left|\rho_1\right|}{\left|\rho_2\right|}}\varrho_1^{(\text{1-C-1})}\right\}, \nonumber \\
&\left\{\varrho_1^{(\text{1-D})}e^{j\angle{h_1^*h_2}},\sqrt{\frac{\left|\rho_1\right|}{\left|\rho_2\right|}}\varrho_1^{(\text{1-D})}\right\}, 
\nonumber\\& \left\{\varrho_1^{(\text{1-C-1})}e^{j\angle{h_1^*h_2}+j\pi},\sqrt{\frac{\left|\rho_1\right|}{\left|\rho_2\right|}}\varrho_1^{(\text{1-C-1})}\right\}, \nonumber\\& 
\left\{\varrho_1^{(\text{2})}e^{j\angle{h_1^*h_2}+j\pi},\sqrt{\frac{\left|\rho_1\right|}{\left|\rho_2\right|}}\varrho_1^{(\text{2})}\right\}\Bigg\}, \label{C-set}
\end{align}
where 
\begin{itemize}
	\item $\varrho_2^{(\text{1-B-1})}=\sqrt{\frac{1}{-2\rho_2}}$ and $\varrho_1^{(\text{1-C-1})}=\sqrt{\frac{1}{-2\rho_1}}$.
	\item  $\varrho_2^{(\text{1-B-2})}$ is the unique positive root of the sixth-order polynomial
	\begin{align}
	2\left|\tilde{h}_2\right|^2\varrho_2^6-6\rho_2\sigma^2\varrho_2^2-\sigma^2. \label{case1-b-2}
	\end{align}
	\item $\varrho_1^{(\text{1-C-2})}$ is the unique positive root of the sixth-order polynomial
	   \vspace{-2mm}
	\begin{align}
	2\left|\tilde{h}_1\right|^2\varrho_1^6-6\rho_1\sigma^2\varrho_1^2-\sigma^2. \label{case1-c-2}
	\end{align}
	\item $\varrho_1^{(\text{1-D})}$ is the unique positive root of the sixth-order polynomial
	\begin{align}
	2\left(\left|\tilde{h}_1\right|+\frac{\left|\rho_1\right|^{3/2}}{\left|\rho_2\right|^{3/2}}\left|\tilde{h}_2\right|\right)^2\varrho_1^6-6\rho_1\sigma^2\varrho_1^2-\sigma^2. \label{case1-d}
	\end{align}
	\item $\varrho_1^{(\text{2})}$ is the only positive root of the sixth-order polynomial 
	\begin{align}
	2\left(\left|\tilde{h}_1\right|-\frac{\left|\rho_1\right|^{3/2}}{\left|\rho_2\right|^{3/2}}\left|\tilde{h}_2\right|\right)^2\varrho_1^6-6\rho_1\sigma^2\varrho_1^2-\sigma^2. \label{case2}
	\end{align}
	
\end{itemize}
\end{thm}
\begin{IEEEproof}
	The proof is provided in Appendix~\ref{appthe2}.
	\end{IEEEproof}

Note that both the characteristics of the power amplifier non-linearity and the backward crosstalk should be known at the transmitter to implement the optimal precoder presented in Theorem~\ref{thm:optimal-precoding}. To determine how important it is to obtain these characteristics, in the following sections, we will consider two sub-optimal precoding vectors that neglect all or some of  the hardware characteristics. In addition, we will also derive the optimal input reference power for these precoders.

\subsection{Conventional MRT \label{conv-mrt}}

In the absence of power amplifier non-linearity and backward crosstalk, conventional MRT is the optimal precoder and it is desirable to transmit with as high power as possible. If we consider MRT in the presence of non-linearities and crosstalk, the distortion and noise in the denominator in \eqref{sndr} also depends on the transmit power and therefore the SNDR is maximized at a finite reference power $P_x$.
In this section, we set
\begin{equation}
{\bf c}=\sqrt{\widetilde{P}_x}{\bf h}^*
\end{equation}
and optimize the power control coefficient $\widetilde{P}_x$ in order to maximize the SE and, equivalently, maximizing the SNDR in \eqref{sndr}. The relation between the power control coefficient $\widetilde{P}_x$ and the actual input reference power $P_x$ is $P_x=\widetilde{P}_x\left|h_1\right|^2$ according to \eqref{Cx2}. The effective precoding vector defined in the previous section is given as ${\bf \tilde{c}}={\bf Q}{\bf c}=\sqrt{P_x}\frac{{\bf Q}{\bf h}^*}{\left|h_1\right|}$. Let us define the fixed part of the effective precoder as ${\bf \hat{c}}\DEF\frac{{\bf Q}{\bf h}^*}{\left|h_1\right|}$, hence ${\bf \tilde{c}}=\sqrt{P_x}{\bf \hat{c}}$. Using the expressions in \eqref{Arate}, the SE maximization problem in terms of $P_x$ can be expressed as
\begin{align} \label{rate_max_con_mrt}
&\underset{P_x}{\text{maximize}} \ \ \ \frac{2P_x\left(\left|k_1\right|^2P_x^2+2\Re\left\{k_0k_1^*\right\}P_x+\left|k_0\right|^2 \right)}{\left|k_1\right|^2P_x^3+\sigma^2}
\end{align}
where the two constants were defined for ease of notation:
\begin{align}
&k_0=h_1\hat{c}_1+h_2\hat{c}_2 \in \mathbb{C}, \ \ \  k_1=\tilde{h}_1|\hat{c}_1|^2\hat{c}_1+\tilde{h}_2|\hat{c}_2|^2\hat{c}_2 \in \mathbb{C}, \label{k1}
\end{align}
where $\tilde{h}_{\ell}$, $\ell=1,2$, and $\sigma^2$ are as in \eqref{htilde}. For conventional MRT, $k_0$ and $k_1$ are not phase aligned since $\angle{\hat{c}_\ell}$ may not be equal to  $\angle{h_\ell^*}$ due to the effect of backward crosstalk. However, $k_0$ and $k_1$ are phase aligned for the optimal precoder as can be seen from Appendix~\ref{appthe2}. Hence, neglecting backward crosstalk yields some drop in the SE.

To solve the one-dimensional optimization problem in \eqref{rate_max_con_mrt}, we take the derivative of the objective function and equate it to zero. We then obtain the candidate solutions as the positive roots of the following fourth-order polynomial of $P_x$:
\begin{align} \label{derivative_con_mrt}
&2\left|k_1\right|^2\Re\left\{k_0k_1^*\right\}P_x^4+2\left|k_1\right|^2\left|k_0\right|^2P_x^3\nonumber\\&-3\left|k_1\right|^2\sigma^2P_x^2-4\Re\left\{k_0k_1^*\right\}\sigma^2P_x-\left|k_0\right|^2\sigma^2.
\end{align}
The optimal input reference power for $P_x$ is the root that maximizes the SNDR in \eqref{rate_max_con_mrt}.

{\bf Remark:} Note that we have not been put any constraint on $P_x$ in any of the optimization problems considered so far. However, the Bussgang gains $a_\ell$ are greater than zero in practice meaning that $1+2\rho_\ell\left|\hat{c}_\ell\right|^2P_x>0$, for $\ell=1,2$. However, $\rho_\ell<0$ is usually very small in absolute value compared to 1 and only very large input power can make the Bussgang gains negative. Hence, we have not considered these practically implicit constraints. We also note that as $P_x\to\infty$, it is clearly seen that the SNDR in \eqref{rate_max_con_mrt} approaches 2. Since we consider practical range of input reference powers, we are not interested in this asymptotic behavior and evaluate only the critical points of the objective function.

\subsection{Distortion-Aware MRT \label{dist-aw-mrt}}

In this section, we will find the optimal input power for another sub-optimal precoder which selects ${\bf \tilde{c}}={\bf Q}{\bf c}=\sqrt{\eta}{\bf A}^T{\bf h}^*$ in order to maximize the desired signal strength in the numerator of the SNDR. We call this precoder distortion-aware MRT and note that  $\eta>0$ is a power control coefficient. Unlike the previous case, we cannot find the optimal input reference power, $P_x$, by optimizing $\eta$ for a fixed value of ${\bf A}^T{\bf h}^*$ since also the Bussgang matrix ${\bf A}$ depends on $P_x$. More precisely, the elements of ${\bf \tilde{c}}$ are given as
\begin{align}
\tilde{c}_\ell=\sqrt{\eta}h_\ell^{*}\left(1+2\rho_\ell\left|\tilde{c}_\ell\right|^2\right).
\end{align} 
Assuming the Bussgang gains $a_\ell=\left(1+2\rho_\ell\left|\tilde{c}_\ell\right|^2\right)$ are positive, the phase of $\tilde{c}_\ell$ is $\angle{h_\ell^{*}}$. However, there is a dependency between their gains as
\begin{align} \label{eta}
& \sqrt{\eta}=\frac{\left|\tilde{c}_1\right|}{\left|h_1\right|\left(1+2\rho_1\left|\tilde{c}_1\right|^2\right)}=\frac{\left|\tilde{c}_2\right|}{\left|h_2\right|\left(1+2\rho_2\left|\tilde{c}_2\right|^2\right)}.
\end{align}
By arranging the terms in \eqref{eta}, we obtain quadratic equations of  $\left|\tilde{c}_1\right|$ and $\left|\tilde{c}_2\right|$, which both have only one positive root. Hence, for a given $\eta$, the elements of the effective precoding vector ${\bf \tilde{c}}$ are given as
\begin{align}\label{eta_c}
& \tilde{c}_\ell=\frac{1-\sqrt{1-8\rho_\ell\left|h_\ell\right|^2\eta}}{4\rho_\ell\left|h_\ell\right|\sqrt{\eta}}e^{j\angle{h_\ell^*}}, \ \ \ell=1,2. 
\end{align}
Due to the one-to-one relationship between $\tilde{c}_1,\tilde{c}_2$ and $\eta$, the SNDR maximization problem in \eqref{rate_max} can be expressed as a one-dimensional optimization problem in terms of $\eta$. However, the resultant objective function is complicated due to the square root expressions and finding its critical points are not easy. Instead, one can simply make a line search over $\eta$ to find the SE and the corresponding input reference power $P_x=\left|c_1\right|^2$ where the actual precoder is given by ${\bf c}={\bf Q}^{-1}{\bf \tilde{c}}$.

\section{Numerical Results and Discussion \label{numerical}}

In this section, we simulate the impact of backward crosstalk and power amplifier non-linearity on the NMSE for a $2\times2$ MIMO transmitter and on the SE of a $2\times1$ MISO channel when using this $2\times2$ transmitter.

\begin{figure}[t!]
		\begin{center}
			\includegraphics[trim={1.8cm 0cm 2.6cm 0.2cm},clip,width=8cm]{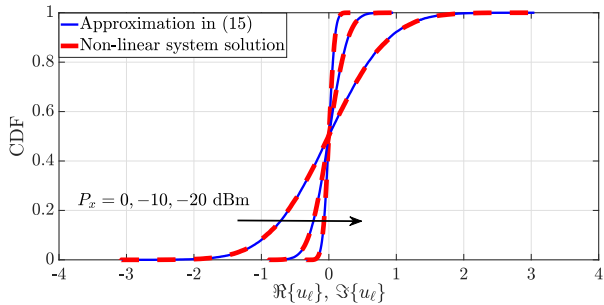}
			\caption{The CDF of $\Re\{u_{\ell}\}$ and $\Im\{u_{\ell}\}$ for using the derived approximation and the solution of the non-linear system.} \label{fig:sim1a}
		\end{center}
\end{figure}

First, we validate the Gaussian approximation derived in \eqref{u_small_errorb} by the solution of the non-linear equation system in \eqref{u} with \eqref{r} for a randomly generated Gaussian input $\bfx$ of length 10000. Then the corresponding $\bfu$ is obtained by using the ``fsolve'' function in MATLAB with the initial point ${\bf Q}\bfx$. In this setup, the input power of two branches are the same ($\beta=1$) and the correlation coefficient is $\xi=0$. We consider a fully symmetric transmitter: $\kappa_1=\kappa_2=\kappa$, $\rho_1=\rho_2=\rho$, and $\gamma_1=\gamma_2=\gamma$. The backward crosstalk parameter is selected as real and positive, and its power is $10\log_{10}(|\kappa|^2)=-50$\,dB. The power amplifier compression parameter is $\rho=-0.025$, and the thermal noise variance at the transmitter is $\sigma_w^2=-10$\,dBm. The gain of power amplifier is $10\log_{10}(\gamma^2)=30$\,dB,\footnote{Note that the power of the leakage signal due to crosstalk is $10\log_{10}(|\kappa|^2)+10\log_{10}(\gamma^2)=-20$\,dB for this setup.} where $\gamma>0$ and $\gamma^2$ are the amplification and power gains, respectively. Since the transmitter is fully symmetric, both $u_1$ and $u_2$ have the same distribution and the cumulative distribution function (CDF) of the real and imaginary parts of them are shown in Fig.~\ref{fig:sim1a} for three values of input reference power: $P_x=-20$, $P_x=-10$, and $P_x=0$\,dBm. As Fig.~\ref{fig:sim1a} shows, the approximation in \eqref{u_small_errorb} matches well with the exact solution of the non-linear system. Furthermore, the NMSE for the covariance of $u_1$ and $u_2$ between two sets of data is given as $-32$\,dB, $-22$\,dB, and $-29$\,dB, respectively for $P_x=-20$, $P_x=-10$, and $P_x=0$\,dBm.

\subsection{NMSE Performance}

\begin{figure}[t!]
	\begin{center}
		\includegraphics[trim={1.8cm 0cm 2.6cm 0.2cm},clip,width=8cm]{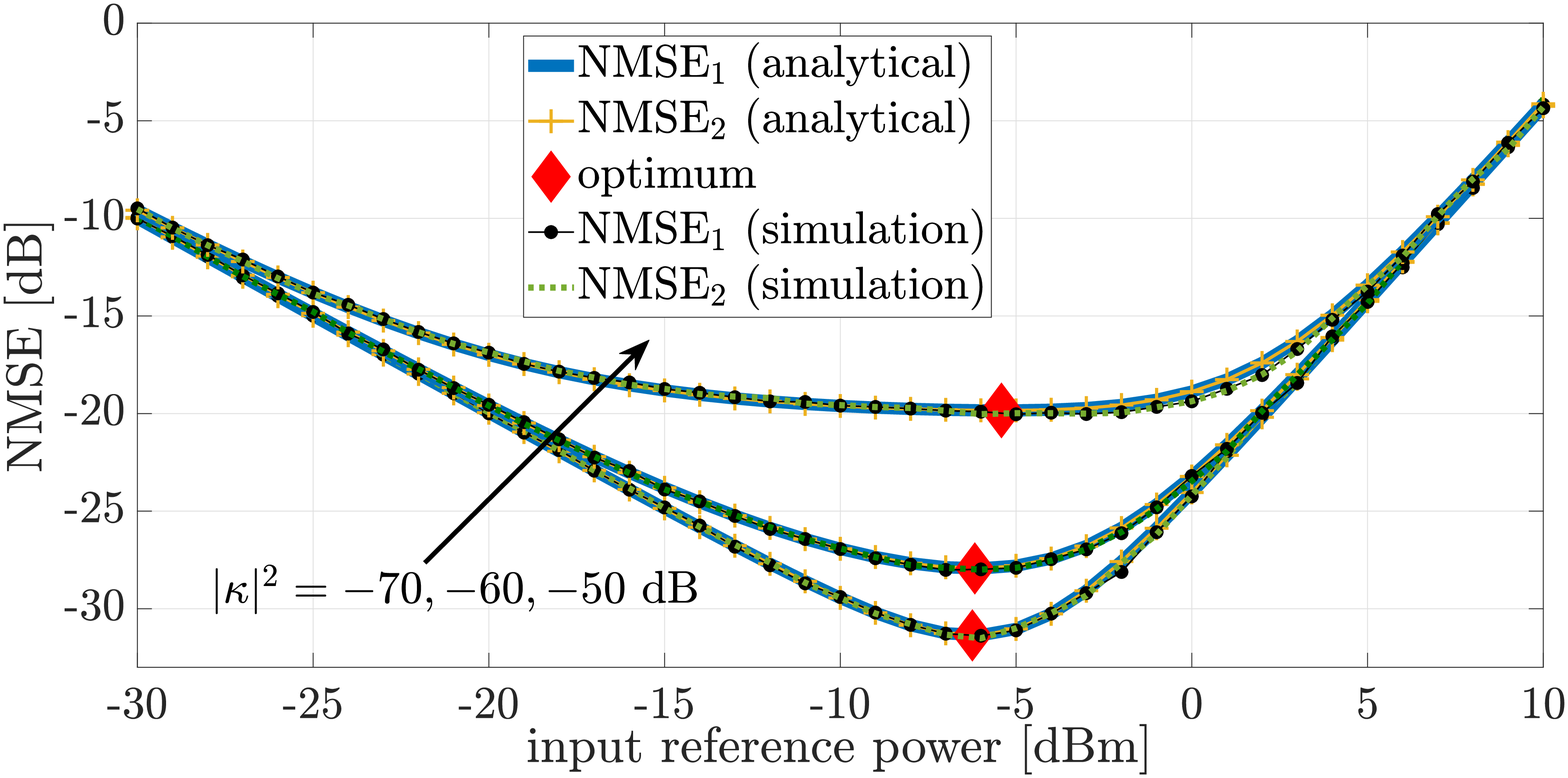}
		\caption{NMSE versus input reference power, $\Px$ for a symmetric transmitter.} \label{fig:sim1}
	\end{center}
\end{figure}

We consider again the fully symmetric transmitter described above. Hence, the NMSE of the two branches are identical:  $\NMSE_1(\Px)=\NMSE_2(\Px)$. Fig.~\ref{fig:sim1} shows $\NMSE=\NMSE_1=\NMSE_2$ versus the input reference power, $\Px$ for three levels of crosstalk: $|\kappa|^2=-70$\,dB, $|\kappa|^2=-60$\,dB, and $|\kappa|^2=-50$\,dB. Both the derived analytical results and simulation results obtained by averaging the Monte Carlo trials with the exact solution of the non-linear system are shown. The optimal solution of the min-max NMSE problem is shown by the red diamond. Since the NMSE for two branches is the same, the optimal $\Px$ is the minimizer of $\NMSE_1(\Px)=\NMSE_2(\Px)$. Although there are some slight deviations between the simulation and the analytical results obtained by the Gaussian approximation for $\bfu$ at some points, the closed-form results match very well with the simulation.

As seen from Fig.~\ref{fig:sim1}, as the crosstalk power decreases, the NMSE becomes more sensitive to the changes in input reference power, $P_x$. For $|\kappa|^2=-70$\,dB, the optimum $P_x$ is around -6\,dBm. When we decrease $P_x$ to -10\,dBm or increase it to -2\,dBm, the NMSE increases by around 2\,dB. Moreover, the optimal $P_x$ also changes with the crosstalk level.

\begin{figure}[t!]
		\begin{center}
			\includegraphics[trim={1.8cm 0cm 0.2cm 0.2cm},clip,width=8.1cm]{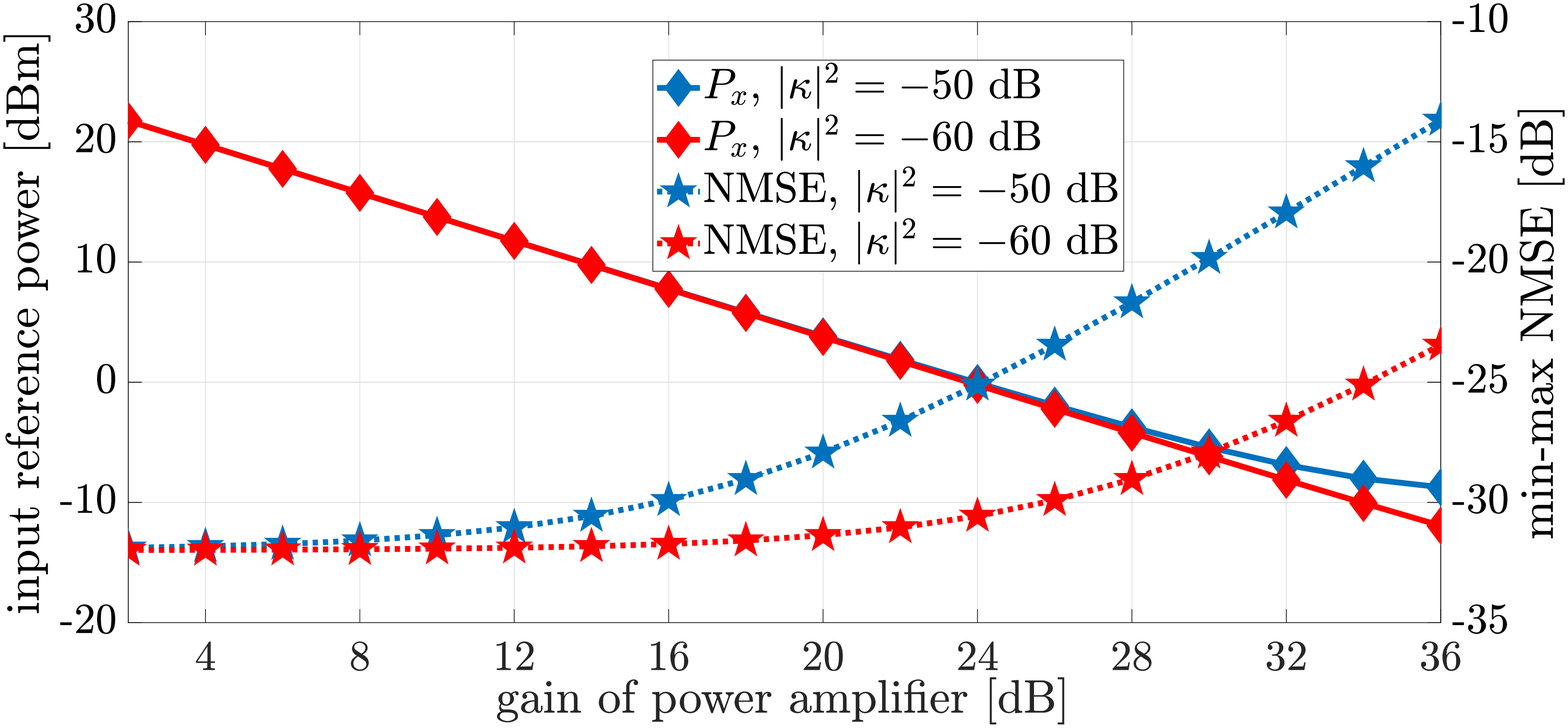}
		\caption{Optimal input reference power, $\Px$ and NMSE versus power gain, $\gamma^2$, for a symmetric transmitter.} \label{fig:sim2}
		\end{center}
\end{figure}

\begin{figure}[t!]
	\begin{center}
		\includegraphics[trim={1.8cm 0cm 2.6cm 0.2cm},clip,width=8cm]{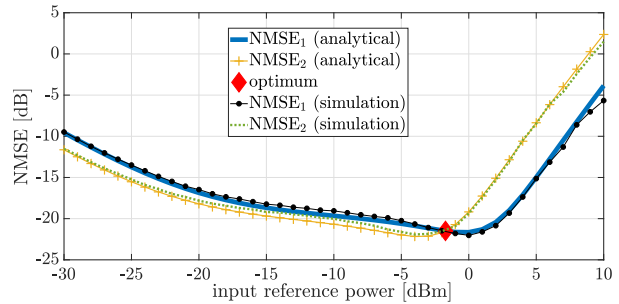}
		\caption{NMSE versus input reference power, $\Px$ for the asymmetric transmitter.} \label{fig:sim3}
	\end{center}
\end{figure}

In Fig.~\ref{fig:sim2}, we plot the optimal input reference power that minimizes $\NMSE_1=\NMSE_2$ and the corresponding NMSE by changing the power gain of amplifier, $10\log_{10}(\gamma^2)$\,dB. We consider two different backward crosstalk parameters: $|\kappa|^2=-50$\,dB or $|\kappa|^2=-60$\,dB. For both cases, as the gain of power amplifier increases, the value of $P_x$ that minimizes the NMSE decreases. For relatively smaller values of $\gamma^2$, optimal input reference power for both crosstalk levels are nearly the same. After some point, the gap between the input powers starts to become visible. Higher input power is needed at the optimal point when the crosstalk level increases. As expected, the corresponding optimal NMSE is higher for $|\kappa|^2=-50$\,dB where the gap increases with the power amplifier gain.

Next, we consider the performance of an asymmetric transmitter with different power levels and non-zero correlation between the two input signals. The amplitude ratio of the second input signal to the first one is taken as  $\beta=1.3$. The correlation coefficient is $\xi=0.7$. The backward crosstalk parameters are $10\log_{10}(|\kappa_1|^2)=-48$\,dB and $10\log_{10}(|\kappa_2|^2)=-52$\,dB. The power amplifier compression parameters are $\rho_1=-0.023$, and $\rho_2=-0.027$. The gains of the power amplifiers are the same: $10\log_{10}(\gamma^2)=30$\,dB.

Fig.~\ref{fig:sim3} shows $\NMSE_1$ and $\NMSE_2$ for the asymmetric transmitter versus the input reference power, $\Px$. The optimal solution of the min-max NMSE problem for this setup is where $\NMSE_1(\Px)=\NMSE_2(\Px)$ which corresponds to Case 3 in Appendix~\ref{appthe1}. Note that $\NMSE_1$ achieves its minimum at a clearly different point. When $P_x$ deviates from its optimal value, either $\NMSE_1$ or $\NMSE_2$ increases and $\max\{ \NMSE_1, \NMSE_2\}$ changes substantially. In addition, the optimal $P_x$ for the exact simulation data is approximately the same with the analytical one. This shows that the derived analytical results model the effect of dirty MIMO transmitter properly.

\begin{figure}[t!]
		\begin{center}
			\includegraphics[trim={1.8cm 0cm 0.3cm 0.2cm},clip,width=8.1cm]{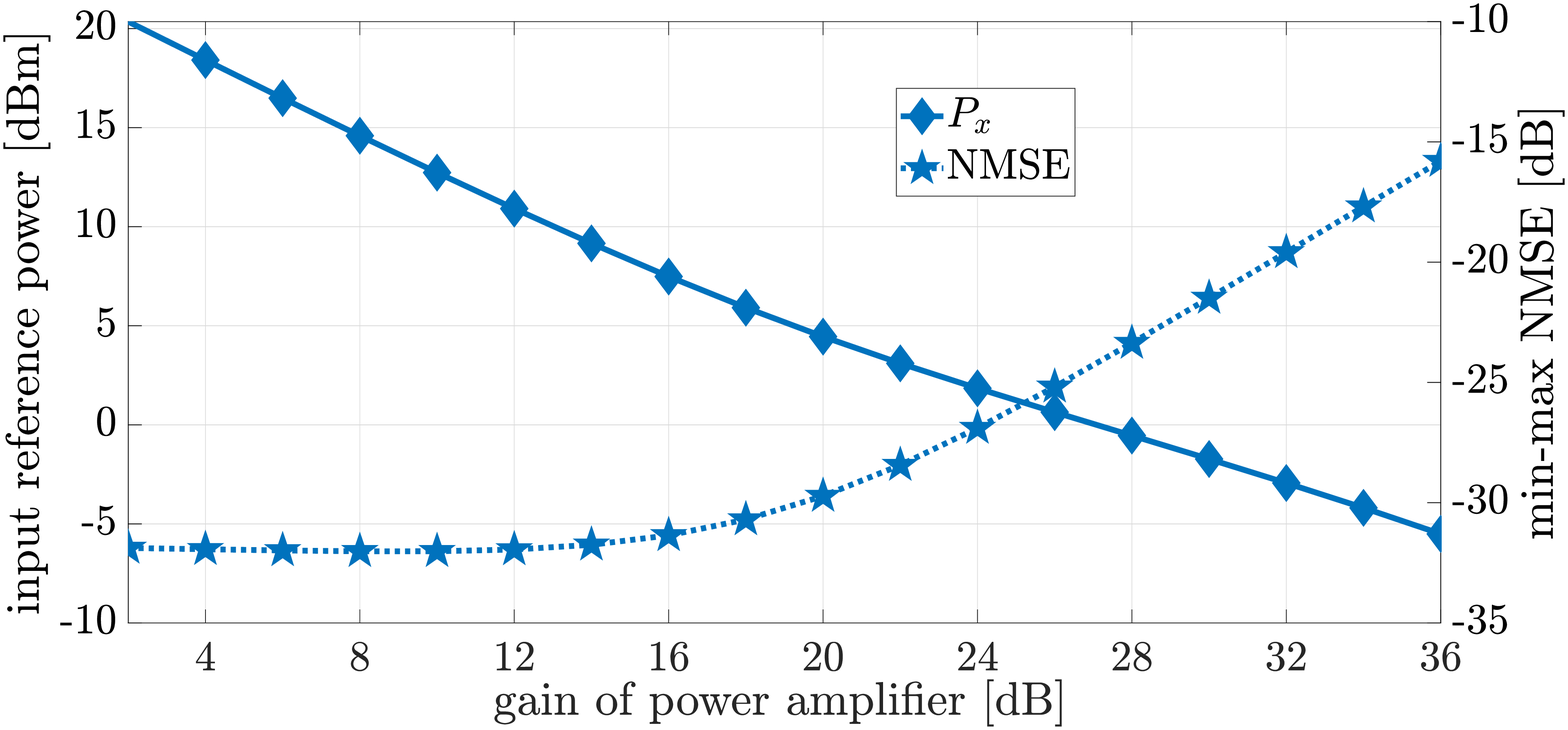}
		\caption{Optimal input reference power, $\Px$ and NMSE versus $\gamma^2$ for the asymmetric transmitter.} \label{fig:sim4}
		\end{center}
\end{figure}

\begin{figure}[t!]
	\begin{center}
		\includegraphics[trim={1.8cm 0cm 2.6cm 0.2cm},clip,width=8cm]{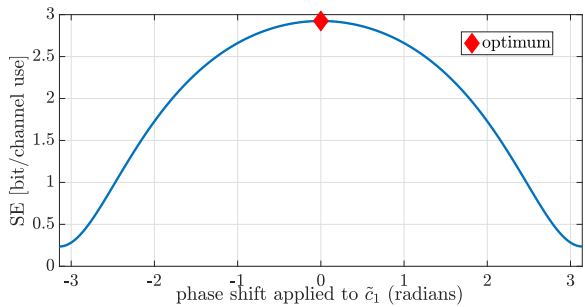}
		\caption{SE versus the phase shift applied to the optimal $\tilde{c}_1$.} \label{fig:sim5}
	\end{center}
\end{figure}

We repeat the experiment in Fig.~\ref{fig:sim2} but in the asymmetric transmitter case with $10\log_{10}(|\kappa_1|^2)=-48$\,dB and $10\log_{10}(|\kappa_2|^2)=-52$\,dB, and show the results in Fig.~\ref{fig:sim4}. We observe similar characteristics, except that the NMSE is around 5\,dB greater for relatively small values of $\gamma^2$.

\subsection{SE Performance}

Now, we will consider the SE performance in the symmetric transmitter case described above. The channel coefficients are randomly and independently generated as $h_1,h_2\sim \mathcal{N}_{\mathbb{C}}(0,1)$. The thermal noise variance at the receiver is taken as $\sigma_n^2$. Hence, channel gain over noise ratio is given by $1/\sigma_n^2$.

\begin{figure}[t!]
		\begin{center}
			\includegraphics[trim={1.8cm 0cm 2.6cm 0.2cm},clip,width=8cm]{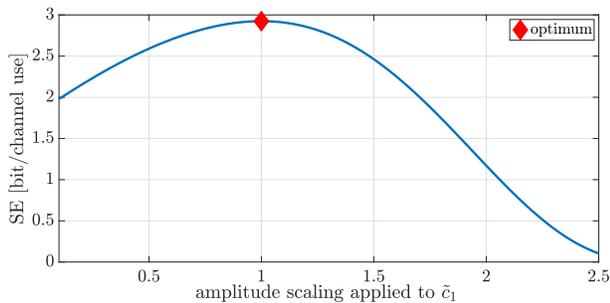}
		\caption{SE versus the amplitude scaling applied to the optimal $\tilde{c}_1$.} \label{fig:sim6}
		\end{center}
\end{figure}

We first consider a single channel realization with $\sigma_n^2=1$ and verify the optimality of the proposed precoder in Theorem~\ref{thm:optimal-precoding} that maximizes SE. Fig.~\ref{fig:sim5} shows the SE  when the first element of the optimal effective precoder, $\tilde{c}_1$, is phase shifted and all other parameters of the optimal precoder are kept constant. The zero phase shift corresponds to the proposed optimal precoder. As it can be seen from Fig.~\ref{fig:sim5}, when the phase of the optimal $\tilde{c}_1$ is shifted, the SE can be reduced significantly compared to the optimal value. 
In Fig.~\ref{fig:sim6}, we are instead keeping the phases of the elements of the optimal effective precoder to be the same and scale $\tilde{c}_1$ to see how the SE changes. We notice that the maximum SE is achieved when the scaling parameter is 1, which verifies the optimality of the proposed SE maximizing precoder. As the amplitude scaling increases above 1, the SE drops quickly towards zero. This is due to the decrease in the amplitude of the corresponding Bussgang gain and increased distortion level.

\begin{figure}[t!]
	\begin{center}
	\includegraphics[trim={1.8cm 0cm 2.6cm 0.2cm},clip,width=8cm]{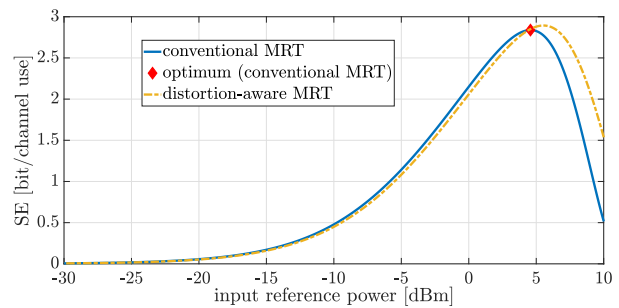} 
	\caption{SE versus input reference power, $\Px$, for conventional and distortion-aware MRT.}  
	\label{fig:sim7}
	\end{center}
\end{figure}

For the same single channel realization, Fig.~\ref{fig:sim7} shows the SE versus input reference power, $P_x$, for conventional and distortion-aware MRT. The proposed optimal solution for conventional MRT is shown by the red diamond and it obviously maximizes the SE. The SE curve for distortion-aware MRT is obtained by a line search over $\eta$ and using the relations in \eqref{eta_c}. We observe that the SE versus $P_x$ has a uni-modal characteristic. Furthermore, the gap between the SE curves for two precoders are very close for $P_x\leq5$\,dBm. For higher values of $P_x$, the distortion-aware MRT provides higher SE compared to the conventional MRT by exploiting the distortion characteristics. We note that although these two sub-optimal precoders do not utilize the full information related to the crosstalk distortion, the optimal $P_x$ is dependent on the all the hardware parameters. Hence, by neglecting crosstalk, the SE may deteriorate significantly. In the following experiments, we will find the optimal $P_x$ for the distortion-aware MRT by a line search over $\eta$.
Interestingly, the SE-maximizing input power is substantially higher (around 10\,dB) than the NMSE-minimizing input power, which was shown in Fig.~\ref{fig:sim1}.

\begin{figure}[t!]
		\begin{center}
			\includegraphics[trim={1.8cm 0cm 2.6cm 0.2cm},clip,width=8cm]{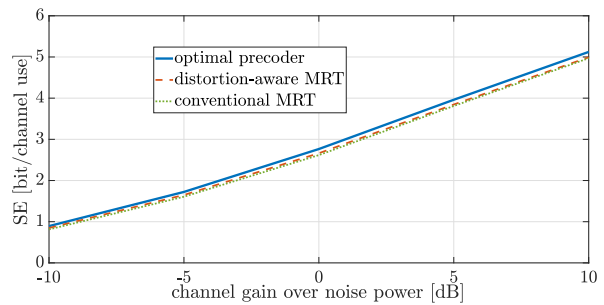}
			\caption{Average SE versus channel gain over noise $1/\sigma_n^2$.} \label{fig:sim9}
		\end{center}
	\end{figure}

We now consider 1000 random channel realizations and plot the average SE for all the considered precoders: a) the optimal precoder that maximizes SE, b) distortion-aware MRT with optimized $P_x$, c) conventional MRT with optimized $P_x$. In Fig.~\ref{fig:sim9}, the SE performance of the three precoders is considered by changing channel gain over noise. Although there is a slight difference between the SE of the considered precoders, it is possible to attain the same SE with approximately 0.7\,dB and 0.5\,dB worse channel gain compared to the conventional MRT and the distortion-aware MRT, respectively. Hence, although the hardware distortion has clear impact on the SE, the same precoding still works fairly well.

	\begin{figure}[t!]
		\begin{center}
			\includegraphics[trim={1.4cm 0cm 2.6cm 0.2cm},clip,width=8.1cm]{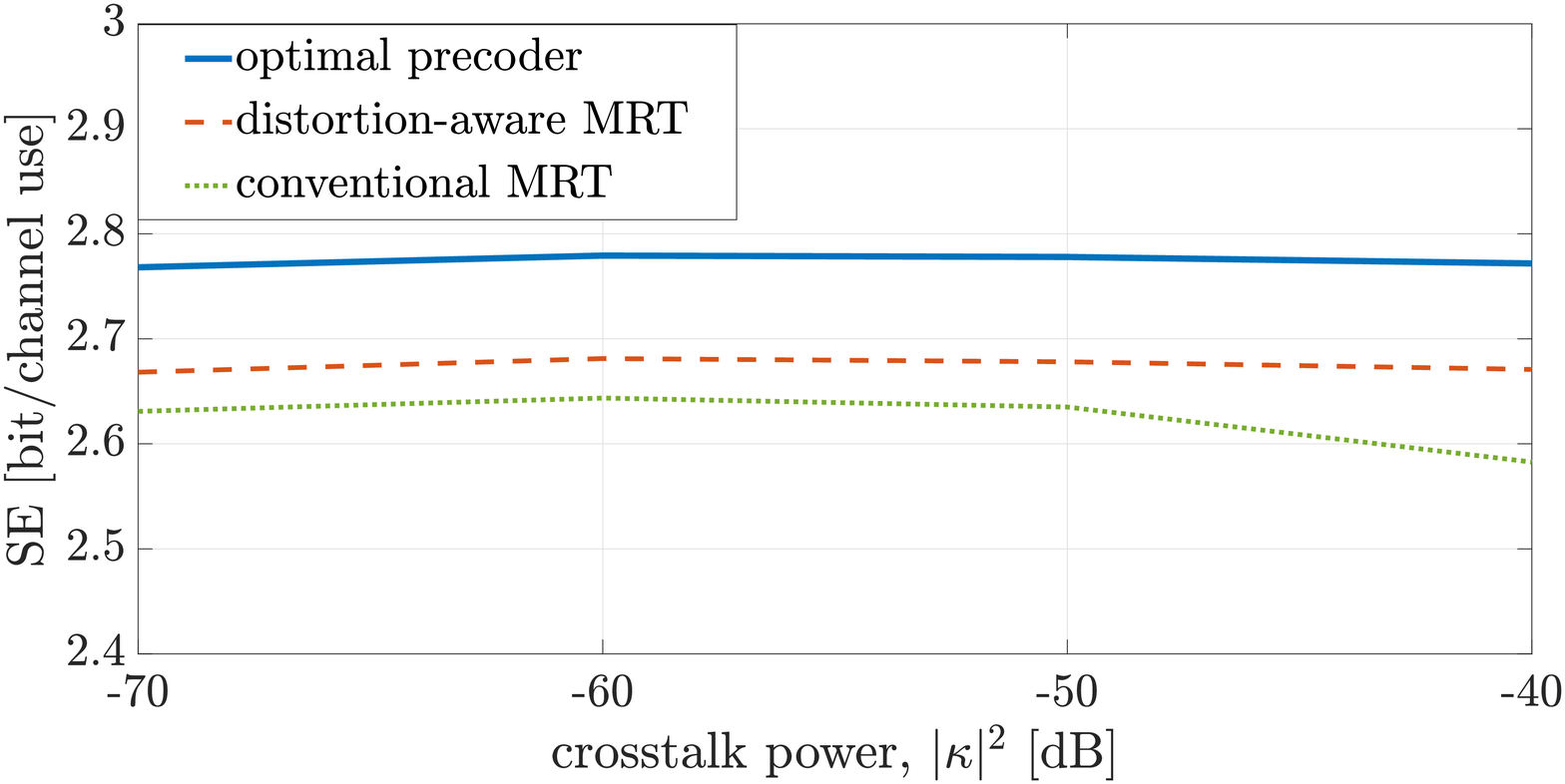}
		\caption{Average SE versus crosstalk power, $|\kappa|^2$.} \label{fig:sim10}
		\end{center}
\end{figure}

In Fig.~\ref{fig:sim10}, we set the channel gain over noise to 0\,dB and analyze the effect of the crosstalk parameter, $\kappa$, on the SE. The performance of the optimal precoder and the distortion-aware MRT is barely affected by the crosstalk change. This is due to the fact that they both exploit the structure of the matrix ${\bf Q}$ that is determined by the power amplifier gains and crosstalk parameters. The non-diagonal structure of this matrix yields that the optimal phase difference between the precoder weights is not equal to the phase difference between the channels of two antennas. In conventional MRT, the effect of ${\bf Q}$ is neglected and the phase deterioration results in a consistent drop in the SE with the increase of crosstalk strength after -60\,dB although the input reference power is also optimized.

According to our observations, the optimal input reference power for maximum SE changes substantially with different channel realizations even if the channel variance is the same. This is typical for fading channels. Furthermore, it may significantly differ from the input reference power that minimizes the maximum NMSE of the two transmitter branches. One can often use a higher power for data transmission since the SE grows with $P_x$ as long as the numerator of the SNDR increases faster than the denominator.

\section{Extension of $M \times M$ MIMO Transmitters \label{extension}}

In this section, we will discuss how the analytical results derived in the previous sections can be extended to $M\times M$ MIMO transmitters. For such a transmitter with $M$ antennas, a feedback connection can be drawn as in Fig.~\ref{figdirty} between two different antenna branches with crosstalk parameters $\kappa_{\ell,m}\in \mathbb{C}$, where $\kappa_{\ell,m}$ is the scaling factor from the $\ell$th antenna branch to the $m$th one. The input signal is $\bfx=(x_1 \ldots x_M)^T \sim \mathcal{N}_{\mathbb{C}}\left({\bf 0}, {\bf C_x}\right)$ where ${\bf C_x} \in \mathbb{C}^{M\times M}$ depends only one design parameter that is $\Px = \E [ |x_1|^2 ]$ as in \eqref{Cx}. The output of the transmitter is given by $\bfy = \bfr + \bfw$, where the only difference compared to Section~\ref{model} is that all the vectors are $M$-dimensional. The power amplifier outputs can be represented as in \eqref{r}  with an $M$-dimensional counterpart of the diagonal matrix ${\bf G}$ and the third-order distortion function $\bff(\bfu)$. All the equations in (\ref{r_Au})--(\ref{ub}) are also valid for $M\times M$ transmitter with appropriately defined matrices. Note that the gain matrix $\bfL$ is diagonal with entries $\gamma_{\ell}$, for $\ell=1,\ldots,M$ and the feedback matrix $\bfK$ has zero diagonal elements with the $\left(\ell,m\right)$th element being $\gamma_{\ell}\kappa_{m,\ell}$, for $\ell\neq m$. Using the same small-error approximations, a similar relationship can be obtained as in \eqref{u_small_errorb}, where the elements of the matrix ${\bf Q}\in \mathbb{C}^{M \times M}$ all depend on the constant system parameters. Using this linear relation, the covariance matrix $\bfU$ becomes $P_x{\bf T}$ where the elements of the constant matrix ${\bf T}\in \mathbb{C}^{M \times M}$, i.e., $t_{\ell m}$ can easily be determined as in (\ref{t11n})--(\ref{t22n}). Furthermore, the Bussgang matrix $\bfA$ is diagonal with elements $a_{\ell}=1+2\rho_{\ell}u_{\ell\ell}$ when  the power amplifier non-linearities affect each antenna signal separately \cite[Sec.~II.B]{Emilny}.

Note that the simplified expressions for the fourth-order and the sixth-order matrices in Appendix~\ref{V_matrix} hold for any $M$ with proper dimensions. Using this fact and taking similar steps as in Section~\ref{nmse}, we can express the diagonal elements of the error covariance matrix $\bfE$ as the third-order polynomial functions of the reference input power, $P_x$ with the coefficient of $P_x^3$ terms being positive. Hence, the NMSE for each transmitter branch becomes a convex function of $P_x$ as claimed by Lemma~\ref{lemma:convex-functions} and the min-max fair optimization problem in \eqref{objective} is a convex programming problem. When $M$ is relatively large and the transmitter is asymmetric, checking all the candidate solutions in Theorem~\ref{thm1} is not efficient and a numerical solver should be utilized. 

The optimal precoder that maximizes the SE may not be obtainable in a simple form for $M>2$ since the number of canditate solutions to be checked increases exponentially. However, for conventional MRT and distortion-aware MRT, the same steps in Section~\ref{conv-mrt} and Section~\ref{dist-aw-mrt} can be followed to obtain the similar one-dimensional optimization problems with properly defined parameters.

\section{Conclusions \label{conclusion}}

In this paper, a non-ideal $2\times 2$ MIMO transmitter subject to backward crosstalk and power amplifier non-linearities  has been analyzed using Bussgang theory for OFDM transmission. 
By utilizing the signal statistics, the feedback model for the backward crosstalk was reformulated as an approximately linear relation between the transmitter outputs and inputs.
The NMSE compared to the ideal amplified signal at the transmitter output was derived in closed form. It was used to find the power back-off that minimizes the maximum NMSE of two branches. In general, the optimal value will not minimize both NMSEs, but find a suitable trade-off.

The SE of transmission to a single-antenna receiver has also been analyzed and a closed-form achievable SE was derived using the fact that the effective distortion noise in the Bussgang decomposition is uncorrelated with the desired communication signal. Three different precoders were considered. The first one maximizes the SE by exploiting full knowledge of the parameters in the backward crosstalk and power amplifier non-linearity models. One of the two sub-optimal precoders uses the optimal precoder structure for ideal hardware and the other one assumes partial knowledge about the backward crosstalk. We optimized the power back-off for maximum SE also for the sub-optimal solutions. Simulation results showed that the sub-optimal precoders achieve almost the same SE as the optimal precoder; thus, it is not of critical importance to estimate the hardware parameters in practice. However, when the strength of the crosstalk increases, the SE achieved by the sub-optimal precoder that assumes ideal hardware got worse compared to the others. Finally, we also noticed that the SE is often maximized when transmitting at a higher power than what is minimizing the NMSE.

\appendices

\section{Derivation of the Elements of ${\bf V}$ \label{V_matrix}}

The elements of the matrix $\bfV$ in \eqref{V} depend on the matrices $\bfUb$ and $\bfUbb$, which contain fourth- and sixth-order moments of the input signals. These matrices will be derived in this appendix, which finally leads to the simplified expression in \eqref{V2}.

\subsubsection*{Fourth-Order Moments $\bfUb$}
For the third-order nonlinearity $\bff(\bfu)$ in (\ref{r}), the fourth-order moments in $\bfUb= \E[\bff(\bfu) \, \bfu^H] $ read
\begin{IEEEeqnarray}{rCl}
\ub_{\ell\ell} &\DEF& \E[u_\ell \, |u_\ell|^2 \, u_\ell^*] = 2 \, u_{\ell\ell}^2, \quad \ell=1,2, \label{ub12}\\
\ub_{k\ell} &\DEF& \E[u_k \, |u_k|^2 \, u_{\ell}^*] = 2 \, u_{kk} \, u_{\ell k}^*,  \quad \ell \neq k, \label{ub22}
\end{IEEEeqnarray}
where $\E[ u_k \, |u_k|^2 \, u_\ell^* ] = 2 \, \E[u_k \, u_k^*] \, \E[ u_k \, u_\ell^* ]$  was  used \cite{Reed}.  The higher-order moments matrix $\bfUb$ is now given by
\begin{equation} \label{AUb}
\bfUb = 2 \underbrace{\left( \begin{array}{cc}
u_{11} & 0 \\ 0 & u_{22}
\end{array} \right)}_{\displaystyle \bfB} \, \underbrace{\left( \begin{array}{cc}
u_{11} & u_{12} \\ u_{12}^* & u_{22}
\end{array} \right)}_{\displaystyle \bfU}.
\end{equation}

\subsubsection*{Sixth-Order Moments $\bfUbb$}
For the third-order nonlinearity $\bff(\bfu)$ in (\ref{r}), the components of $\bfUbb= \E[\bff(\bfu) \, \bff(\bfu)^H] $ read
\begin{IEEEeqnarray}{rCl}
\ubb_{\ell\ell} &\DEF& \E[u_\ell \, |u_\ell|^2 \, u_\ell^* \, |u_\ell|^2] = 6 \, u_{\ell\ell}^3 , \quad \ell=1,2, \label{ubb11} \\
\ubb_{12} &\DEF& \E[u_1 \, |u_1|^2 \, u_2^* \, |u_2|^2] = \E[ (u_1  \, u_2^*)^2 \, (u_1 \, u_2^*)^* ] \nonumber\\&=&  4 u_{12} \, u_{11} \, u_{22} + 2 u_{12} \, |u_{12}|^2 , \label{ubb12} 
\end{IEEEeqnarray}
where the following results was used \cite{Reed}:
\begin{IEEEeqnarray}{rCl}
&&\E[ (u_\ell \, u_k^*)^2 (u_\ell \, u_k^*)^* ]\nonumber\\&&=
4 \E[u_\ell \, u_k^*] \, \E[u_\ell \, u_\ell^*] \, \E[u_k \, u_k^*]   + 2 \E[u_\ell \, u_k^*]^2 \, \E[u_\ell^* \, u_k].\label{hurra}
\end{IEEEeqnarray}
The matrix $\bfUbb$ can now be divided into two terms as
\begin{IEEEeqnarray}{rCl}
\bfUbb &=& \label{AUbb}
4 \!  \left( \! \! \! \begin{array}{cc}
u_{11}^3 &  \! \!  \! \! \! \! u_{12} \, u_{11} \, u_{22} \\ u_{12}^* \, u_{11} \, u_{22} &   u_{22}^3
\end{array} \! \!  \!  \right) \nonumber\\&& \!  + \!
2 \!  \left( \! \!  \!  \begin{array}{cc}
u_{11}^3 &  \! \! \! \! \! \! u_{12} \, |u_{12}|^2 \\ u_{12}^* \, |u_{12}|^2 &   u_{22}^3
\end{array} \! \!   \! \right). 
\label{Ubb}
\end{IEEEeqnarray}
The result (\ref{Ubb}) can be expressed in $\bfB$ and $\bfU$ given in (\ref{AUb}) as
\begin{equation} \label{AUbb2}
\bfUbb = 4 \bfB \, \bfU \, \bfB + 2 \bfC,
\end{equation}
where the matrix $\bfC$ was introduced as
\begin{IEEEeqnarray}{rCl}\label{AC}
\bfC = \left(   \begin{array}{cc}
u_{11}^3 &  u_{12} \, |u_{12}|^2 \\ u_{12}^* \, |u_{12}|^2 &   u_{22}^3
\end{array}   \right) .
\end{IEEEeqnarray}

\subsubsection*{Final simplified expression}
It follows directly from (\ref{AUb}) that
\begin{equation}\label{A2x}
\bfUb \, \bfU^{-1}  = 2 \bfB.
\end{equation}
Furthermore, the matrix product $\bfUb  \, \bfU^{-1} \, \bfUb^H$ is required to evaluate the properties of the distortion noise.
Combining (\ref{A2x}) with (\ref{AUb}) gives
\begin{equation}\label{A3x}
\bfUb  \, \bfU^{-1} \, \bfUb^H = 4 \bfB \, \bfU \, \bfB,
\end{equation}
where the result follows from the fact that $\bfB=\bfB^H$, and that $\bfU = \bfU^H$.
From (\ref{AUbb2}) and (\ref{A3x}) it now directly follows that
\begin{IEEEeqnarray}{rCl} \label{AD}
\bfUbb -  \bfUb \, \bfU^{-1} \, \bfUb^H  = 2 \bfC,
\end{IEEEeqnarray}
where $\bfC$ is given in (\ref{AC}). This leads to simplified expression in \eqref{V2}.

\section{Proof of Theorem 1 \label{appthe1}}

There are three cases to be evaluated to achieve the optimal solution to the problem \eqref{objective} since we minimize the maximum of two convex functions. The optimal input reference power, $\Px$ is either the minimizer of $\NMSE_1$ or $\NMSE_2$, or the one at which the $\NMSE_1 = \NMSE_2$ are equal. These three candidate solutions are given as follows:

{\bf Case 1:} The unique minimizer of $\NMSE_1(\Px)$, which is denoted $\Px^{(1)}$, is a solution to the equation 
\begin{align}
& \frac{12\rho_1^2  \, t_{11}^3}{\gamma_1^2} \, \Px + 4\gamma_2t_{11}\rho_1\left(\gamma_2\beta^2\left|\kappa_2\right|^2+\beta\Re\left\{\kappa_2\xi^*\right\}\right)=\frac{\sigma_w^2}{\gamma_1^2 \, \Px^2}, \label{NMSE1c} 
\end{align}
which is obtained by taking the first derivative of $\NMSE_1(\Px)$ with respect to $\Px$ and equating it to zero. Since $\NMSE_1$ is a convex function of $\Px$, the candidate solution for this case is the only positive root of the third-order polynomial of $\Px$ in \eqref{NMSE1_polyn}.

{\bf Case 2:} Similarly, we find the unique minimizer of $\NMSE_2(\Px)$, which is denoted $\Px^{(2)}$, as a solution to the equation
\begin{align}
& \frac{12\rho_2^2  \, t_{22}^3}{\gamma_2^2\beta^2} \, \Px + \frac{4\gamma_1t_{22}\rho_2}{\beta^2}\left(\gamma_1\left|\kappa_1\right|^2+\beta\Re\left\{\kappa_1\xi\right\}\right)=\frac{\sigma_w^2}{\gamma_2^2 \, \beta^2 \Px^2}, \label{NMSE2c} 
\end{align}
which is obtained by equating the first derivative of $\NMSE_2(\Px)$ to zero. There is a  unique positive root of the third-order polynomial of $\Px$ in \eqref{NMSE2_polyn}.

{\bf Case 3:} In this case, we take the candidate solution which satisfies $\NMSE_1(\Px)=\NMSE_2(\Px)$. These are the positive roots of the third-order polynomial in \eqref{case3}. Let $\Px^{(3)}$ denote the positive root of \eqref{case3} which makes $\NMSE_1=\NMSE_2$ the smallest assuming there exists at least one positive root of \eqref{case3}. If all the roots are non-positive, then we can consider only Case 1 and 2.

Considering all the three cases, the optimal input reference signal power, $\Px$, for the problem (\ref{objective}) is given by the element of the set $\mathcal{S}=\Big\{\Px^{(1)},\Px^{(2)},\Px^{(3)}\Big\}$ that minimizes the objective function $\max\big\{\NMSE_1(\Px), \NMSE_2(\Px)\big\}$.

\section{Proof of Theorem 2 \label{appthe2}}

In order to simplify the optimization in  \eqref{rate_max}, we express the optimization variables $\tilde{c}_1$ and $\tilde{c}_2$ in terms of their amplitudes and phase components as follows:
\begin{align}
&\tilde{c}_1=\varrho_1\chi_1, \ \  \tilde{c}_2=\varrho_2\chi_2, \ \  \varrho_1, \ \varrho_2 \in   \mathbb{R}, \ \  \chi_1, \ \chi_2 \in \mathbb{C}, \label{c_amp_phase} \\
&\varrho_1\geq 0,  \ \ \varrho_2 \geq 0, \ \ |\chi_1|^2=1, \ \ |\chi_2|^2=1 \label{c_amp_phase_constraints}.
\end{align}
By neglecting the constant 2 in the numerator of \eqref{rate_max}, we can cast the problem in terms of the amplitude and phase shift variables in \eqref{c_amp_phase}:
\begin{align} 
&\underset{\varrho_1, \ \varrho_2, \, \chi_1, \ \chi_2}{\text{maximize}} \ \ \ \frac{\left| \left(h_1\varrho_1+\tilde{h}_1\varrho_1^3\right)\chi_1+\left(h_2\varrho_2+\tilde{h}_2\varrho_2^3\right)\chi_2 \right|^2}{\left|\tilde{h}_1\varrho_1^3\chi_1+\tilde{h}_2\varrho_2^3\chi_2\right|^2+\sigma^2} \label{rate_max2} \\
& \text{subject to} \ \ \varrho_1\geq 0, \ \  \varrho_2\geq 0, \ \ |\chi_1|^2=1, \ \ |\chi_2|^2=1. \label{rate_constraint2}
\end{align}
We observe that the objective function does not change if a common phase rotation is applied to $\chi_1$ and $\chi_2$.
This means that if $\left\{\chi_1^{\star}, \ \chi_2^{\star}\right\}$ are the optimal phase shifts, then $\left\{\chi_1^{\star}e^{j\theta}, \ \chi_2^{\star}e^{j\theta}\right\}$ result in the same optimal objective for any $\theta \in [0,2\pi)$. Using this observation, we can simply take $\chi_2=1$ and consider the optimization problem in terms of $\varrho_1,\ \varrho_2$, and $\chi_1$. For this reduced-size problem, we obtain the necessary optimality condition for $\chi_1$ as follows:
\begin{align}
&\text{DEN}\left(h_1\varrho_1+\tilde{h}_1\varrho_1^3\right)^{*}\left(h_2\varrho_2+\tilde{h}_2\varrho_2^3\right)\nonumber\\&-\text{NUM}\left(\tilde{h}_1\varrho_1^3\right)^{*}\tilde{h}_2\varrho_2^3=\mathcal{L}_1\chi_1,
\end{align}
where $\text{DEN}$ and  $\text{NUM}$ denote the denominator and numerator of the objective function in \eqref{rate_max2}. Note that they are both real and non-negative. $\mathcal{L}_1$ is the real scaled Lagrange multiplier corresponding to the equality  $|\chi_1|^2=1$. Now, we have
\begin{align}
\chi_1=e^{j\angle\left(\frac{\text{DEN}}{\mathcal{L}_1}\left(h_1\varrho_1+\tilde{h}_1\varrho_1^3\right)^{*}\left(h_2\varrho_2+\tilde{h}_2\varrho_2^3\right)-\frac{\text{NUM}}{\mathcal{L}_1}\left(\tilde{h}_1\varrho_1^3\right)^{*}\tilde{h}_2\varrho_2^3\right)}.
\end{align}
Using the definitions of $\tilde{h}_1$ and $\tilde{h}_2$ in \eqref{htilde} and noting that $\rho_1<0$ and $\rho_2<0$ are real scalars, we see that the angle of $\left(h_1\varrho_1+\tilde{h}_1\varrho_1^3\right)$ is either $\angle{h_1}$ or $\angle{h_1}+\pi$. Similar reasoning applies for $\left(h_2\varrho_2+\tilde{h}_2\varrho_2^3\right)$. Hence, considering the two possible sign values for $\mathcal{L}_1$, we have two possibilities for the angle of $\chi_1$ which are
\begin{align}
\angle{\chi_1}=\angle{h_1^*h_2}, \ \ \text{or} \ \ \angle{\chi_1}=\angle{h_1^*h_2}+\pi. \label{chi1}
\end{align}
Now, our aim is to find the optimal set of $\{\varrho_1,\varrho_2\}$ for these two possibilities. Let us explore two cases one by one.

{\bf Case 1:} In this case, we take the candidate solution $\chi_1=e^{j\angle{h_1^*h_2}}$ together with $\chi_2=1$. The maximization problem in (\ref{rate_max2})-(\ref{rate_constraint2}) can be expressed in terms of $\varrho_1$ and $\varrho_2$ as follows:
\begin{align} 
&\underset{\varrho_1,\varrho_2}{\text{maximize}} \ \ \ \frac{\left(\left|h_1\right|\left(\varrho_1+2\rho_1\varrho_1^3\right)+\left|h_2\right|\left(\varrho_2+2\rho_2\varrho_2^3\right) \right)^2}{\left(\left|\tilde{h}_1\right|\varrho_1^3+\left|\tilde{h}_2\right|\varrho_2^3\right)^2+\sigma^2} \label{rate_max_case1} \\
& \text{subject to} \ \ \varrho_1\geq 0, \ \ \varrho_2\geq 0. \label{rate_constraint_case1}
\end{align}
By KKT conditions for the problem in (\ref{rate_max_case1})-(\ref{rate_constraint_case1}), there are four subcases according to complementary slackness conditions and the Lagrange multipliers for \eqref{rate_constraint_case1}. These cases are as follows:

{\bf Case 1-A:} The first case is $\varrho_1=\varrho_2=0$ where both variables are at the boundary. This case results in zero SNDR and is obviously not the optimal solution.

{\bf Case 1-B:} In this case, we take $\varrho_1=0$ and do not put any constraint on $\varrho_2$. After inserting $\varrho_1=0$ into the objective function and by equating the  derivative of it  with respect to $\varrho_2$ to zero, we obtain
\begin{align}
-\frac{2\varrho_2\left(2\rho_2\varrho_2^2+1\right)\left(2\left|\tilde{h}_2\right|^2\varrho_2^6-6\rho_2\sigma^2\varrho_2^2-\sigma^2\right)}{\left(\left|\tilde{h}_2\right|^2\varrho_2^6+\sigma^2\right)^2}=0,
\end{align}
where we obtain the critical points $\varrho_2^{(\text{1-B-1})}=\sqrt{\frac{1}{-2\rho_2}}$ and the only positive root, $\varrho_2^{(\text{1-B-2})}$, of the polynomial in \eqref{case1-b-2}.
We note the solution sets $\{0, \varrho_2^{(\text{1-B-1})}\}$ and $\{0, \varrho_2^{(\text{1-B-2})}\}$ as candidate optimum.

{\bf Case 1-C:} This case is similar to Case 1-B, by the symmetric structure of the objective, i.e., now we take $\varrho_2=0$ and do not put any constraint on $\varrho_1$. After inserting $\varrho_2=0$ into the objective function and by equating the  derivative of it  with respect to $\varrho_1$ to zero, we obtain the critical points as $\varrho_1^{(\text{1-C-1})}=\sqrt{\frac{1}{-2\rho_1}}$ and the only positive root, $\varrho_1^{(\text{1-C-2})}$, of the polynomial in \eqref{case1-c-2}. We note the solution sets $\{\varrho_1^{(\text{1-C-1})},0\}$ and $\{ \varrho_1^{(\text{1-C-2})},0\}$ as candidate optimum.

{\bf Case 1-D:} In this case, we do not put any constraint on $\varrho_1$ and $\varrho_2$ and equate the Lagrange multipliers corresponding to the inequality constraints to zero.  Then, by equating the  derivatives of the objective function  with respect to $\varrho_1$ and $\varrho_2$ to zero, we obtain the necessary optimality conditions. However, interpreting the resultant equations is hard. Instead, we consider an equivalent optimization problem:
\begin{align} 
&\underset{\varrho_1,\ \varrho_2, \ \tilde{\varrho}_1, \ \tilde{\varrho}_2}{\text{maximize}} \ \ \ \frac{\left(\left|h_1\right|\varrho_1-\left|\tilde{h}_1\right|\tilde{\varrho}_1+\left|h_2\right|\varrho_2-\left|\tilde{h}_2\right|\tilde{\varrho}_2\right)^2}{\left(\left|\tilde{h}_1\right|\tilde{\varrho}_1+\left|\tilde{h}_2\right|\tilde{\varrho}_2\right)^2+\sigma^2} \label{rate_max_case1-D} \\
& \text{subject to} \ \ \tilde{\varrho}_1=\varrho_1^3, \ \ \tilde{\varrho}_2=\varrho_2^3. \label{rate_constraint_case1-D}
\end{align}
Note that we have included additional auxiliary variables, $\tilde{\varrho}_1=\varrho_1^3$ and  $\tilde{\varrho}_2=\varrho_2^3$ in order to simplify the KKT conditions and omitted the non-negativity constraints for $\varrho_1$ and $\varrho_2$ since we assume in this case (Case 1-D), the Lagrange multipliers corresponding to these constraints are zero. However, we should select only the non-negative solutions of the problem (\ref{rate_max_case1-D})-(\ref{rate_constraint_case1-D}) to be a candidate for the original problem. KKT conditions for the above problem are given as follows:
\begin{align}
&\frac{2\left|h_\ell\right|\text{NUM}}{\text{DEN}}=3\mathcal{L}_\ell\varrho_\ell^2, \ \ \ \ell=1,2, \label{case1-d-kkt-1} \\
&-2\left|\tilde{h}_\ell\right|\left(\frac{\text{NUM}}{\text{DEN}}+\frac{\text{NUM}^2\left(\left|\tilde{h}_1\right|\tilde{\varrho}_1+\left|\tilde{h}_2\right|\tilde{\varrho}_2\right)}{\text{DEN}^2}\right)=-\mathcal{L}_\ell,  \nonumber\\& \ell=1,2, \label{case1-d-kkt-3} 
\end{align}
where $\mathcal{L}_1$ and $\mathcal{L}_2$ are the Lagrange multipliers corresponding to the equalities in \eqref{rate_constraint_case1-D}. $\text{NUM}$ and $\text{DEN}$ denote the term whose square is taken in the numerator and the denominator term of the objective function in \eqref{rate_max_case1-D}, respectively, for ease of notation. By dividing both sides of \eqref{case1-d-kkt-3} for $\ell=1$ and $\ell=2$, we obtain $\mathcal{L}_1/\mathcal{L}_2=\left|\tilde{h}_1\right|/\left|\tilde{h}_2\right|$. Using this relation in (\ref{case1-d-kkt-1}), we have
\begin{align}
\varrho_2=\sqrt{\frac{\left|\rho_1\right|}{\left|\rho_2\right|}}\varrho_1. \label{case-1-d-relation}
\end{align}
If we insert $\varrho_2$ in \eqref{case-1-d-relation} into the original objective function, we obtain a similar problem as in Case 1-C and the critical points are  $\varrho_1^{(\text{1-C-1})}=\sqrt{\frac{1}{-2\rho_1}}$ and the  only positive root, $\varrho_1^{(\text{1-D})}$, of the polynomial in \eqref{case1-d}. We note the solution sets $\{\varrho_1^{(\text{1-C-1})},\sqrt{\frac{\left|\rho_1\right|}{\left|\rho_2\right|}}\varrho_1^{(\text{1-C-1})}\}$ and $\{\varrho_1^{(\text{1-D})},\sqrt{\frac{\left|\rho_1\right|}{\left|\rho_2\right|}}\varrho_1^{(\text{1-D})}\}$ as candidate optimum. 

{\bf Case 2:} In this case, we take the other candidate solution $\chi_1=e^{j\angle{h_1^*h_2}+j\pi}$ in \eqref{chi1} together with $\chi_2=1$. In this case, the SNDR maximization problem in (\ref{rate_max2})-(\ref{rate_constraint2}) can be expressed in terms of variables $\varrho_1$ and $\varrho_2$ as follows:
\begin{align} 
&\underset{\varrho_1, \ \varrho_2}{\text{maximize}} \ \ \ \frac{\left(\left|h_1\right|\left(\varrho_1+2\rho_1\varrho_1^3\right)-\left|h_2\right|\left(\varrho_2+2\rho_2\varrho_2^3\right) \right)^2}{\left(\left|\tilde{h}_1\right|\varrho_1^3-\left|\tilde{h}_2\right|\varrho_2^3\right)^2+\sigma^2} \label{rate_max_case2} \\
& \text{subject to} \ \ \varrho_1\geq 0, \ \ \varrho_2\geq 0. \label{rate_constraint_case2}
\end{align}
Note that if at least one of $\varrho_1$ or $\varrho_2$ is zero, we obtain the same problem as in Case 1. The only difference occurs when $\varrho_1>0$ and $\varrho_2>0$. In this case, we can follow the same approach in the reformulation (\ref{rate_max_case1-D})-(\ref{rate_constraint_case1-D}). After some straightforward calculations, we obtain the candidate set of solutions: 
$\Big\{\varrho_1^{(\text{1-C-1})},\sqrt{\frac{\left|\rho_1\right|}{\left|\rho_2\right|}}\varrho_1^{(\text{1-C-1})}\Big\}$ and $\Big\{\varrho_1^{(\text{2})},\sqrt{\frac{\left|\rho_1\right|}{\left|\rho_2\right|}}\varrho_1^{(\text{2})}\Big\}$, where $\varrho_1^{(\text{2})}$ is the only positive root of the polynomial in \eqref{case2}.

Now, considering all the candidate solutions, we obtain the optimal $\tilde{c}_1$ and $\tilde{c}_2$ by selecting the one that maximizes the SE, as formulated in \eqref{eq:SE-maximization}. Given the optimal ${\bf \tilde{c}}^{\mathrm{opt}}$, the optimal precoder is given by ${\bf c}={\bf Q}^{-1} {\bf \tilde{c}}^{\mathrm{opt}} $.

\bibliographystyle{IEEEtran}
\bibliography{refs}

\end{document}